\documentclass[12pt,preprint]{aastex}
\usepackage{multirow}
\shorttitle{The M31 Velocity Vector II}
\shortauthors{van der Marel et al.}

\newcommand{\etal}{{et al.~}}

\newcommand{\kms}{\>{\rm km}\,{\rm s}^{-1}}

\newcommand{\Gyr}{\>{\rm Gyr}}
\newcommand{\kpc}{\>{\rm kpc}}
\newcommand{\Mpc}{\>{\rm Mpc}}
\newcommand{\Msun}{\>{\rm M_{\odot}}}

\begin{document}

\title{The M31 Velocity Vector.\\ 
II. Radial Orbit Towards the Milky Way\\ 
and Implied Local Group Mass}

\author{Roeland P.~van der Marel}
\affil{Space Telescope Science Institute, 3700 San Martin Drive, 
       Baltimore, MD 21218}

\author{Mark Fardal}
\affil{Department of Astronomy, University of Massachusetts, Amherst,
       MA 01003}

\author{Gurtina Besla}
\affil{Department of Astronomy, Columbia University, New York, NY 10027}

\author{Rachael L.~Beaton}
\affil{Department of Astronomy, University of Virginia, PO Box 3818, 
       Charlottesville, VA 22903, USA}

\author{Sangmo Tony Sohn, Jay Anderson, Tom Brown}
\affil{Space Telescope Science Institute, 3700 San Martin Drive, 
       Baltimore, MD 21218}

\author{Puragra Guhathakurta}
\affil{UCO/Lick Observatory, Department of Astronomy and Astrophysics, 
       University of California at Santa Cruz, 1156 High Street,
       Santa Cruz, CA 95064}

%%%%%%%%%%%%%%%
% Abstract
%%%%%%%%%%%%%%%

\newpage

\begin{abstract}
We determine the velocity vector of M31 with respect to the Milky Way
and use this to constrain the mass of the Local Group, based on Hubble
Space Telescope proper-motion measurements of three fields presented
in Paper~I. We construct $N$-body models for M31 to correct the
measurements for the contributions from stellar motions internal to
M31. This yields an unbiased estimate for the M31 center-of-mass
motion. We also estimate the center-of-mass motion independently,
using the kinematics of satellite galaxies of M31 and the Local Group,
following previous work but with an expanded satellite sample. All
estimates are mutually consistent, and imply a weighted average M31
heliocentric transverse velocity of $(v_W,v_N) = (-125.2 \pm 30.8 ,
-73.8 \pm 28.4) \kms$. We correct for the reflex motion of the Sun
using the most recent insights into the solar motion within the Milky
Way, which imply a larger azimuthal velocity than previously believed.
This implies a radial velocity of M31 with respect to the
Milky Way of $V_{\rm rad,M31} = -109.3 \pm 4.4 \kms$, and a tangential
velocity $V_{\rm tan,M31} = 17.0 \kms$, with 1$\sigma$ confidence
region $V_{\rm tan,M31} \leq 34.3 \kms$. {\it Hence, the velocity
vector of M31 is statistically consistent with a radial (head-on
collision) orbit towards the Milky Way.} We revise prior estimates for
the Local Group timing mass, including corrections for cosmic bias and
scatter, and obtain $M_{\rm LG} \equiv M_{\rm MW, vir} + M_{\rm M31, vir} =
(4.93 \pm 1.63) \times 10^{12} \Msun$. Summing known estimates for the
individual masses of M31 and the Milky Way obtained from other
dynamical methods yields smaller uncertainties. Bayesian combination
of the different estimates demonstrates that the timing argument has
too much (cosmic) scatter to help much in reducing uncertainties on
the Local Group mass, but its inclusion does tend to increase other
estimates by $\sim 10$\%. We derive a final estimate for the Local
Group mass from literature and new considerations of $M_{\rm LG} =
(3.17 \pm 0.57) \times 10^{12} \Msun$. The velocity and mass results
imply at 95\% confidence that M33 is bound to M31, consistent with
expectation from observed tidal deformations.
\end{abstract}

%%%%%%%%%%%%%%%
% Keywords
%%%%%%%%%%%%%%%


%%%%%%%%%%%%%%%
% Beginning of main text
%%%%%%%%%%%%%%%

\section{Introduction}
\label{s:intro}

The Milky Way (MW) is a member of a small group of galaxies called the
Local Group (LG). The LG is dominated by its two largest galaxies, the
MW and the Andromeda galaxy (M31). The mass and dynamics of this group
have been the topic of many previous studies (e.g., van den Bergh
2000; van der Marel \& Guhathakurta 2008, hereafter vdMG08; Li \&
White 2008; Cox \& Loeb 2008; and references therein). Analysis of
these topics is important for interpretation of structures inside the
LG, such as dark halos, satellite galaxies, and tidal streams. It is
also important for understanding the LG in a proper cosmological
context, since it provides the nearest example of both Large Scale
Structure and hierarchical galaxy formation. While much progress has
been made in understanding the LG mass and dynamics, this has not been
based on actual knowledge of the three-dimensional velocity vector of
M31. This is because until now, the proper motion (PM) of M31 has been
too small to measure with available techniques.

In Paper~I (Sohn, Anderson \& van der Marel 2012) we reported the very
first absolute PMs of M31 stars in three different fields observed
with the Hubble Space Telescope (HST): a field along the minor axis
sampling primarily the M31 spheroid (the ``spheroid field''), a field
along the major axis sampling primarily the M31 outer disk (the ``disk
field''), and a field along M31's Giant Southern Stream (GSS) sampling
primarily the stars that constitute this stream (the ``stream
field''). For each field we measured the average PM of the M31 stars
with respect to the stationary reference frame of background
galaxies. The results are listed in Table~\ref{t:intkin}. PMs in
mas/yr were converted to velocities $(v_W, v_N)$ in km/s in the
directions of West and North using the known distance $D$ of
M31. Throughout this paper we adopt $D = 770 \pm 40 \kpc$ (see
references in vdMG08). The velocity uncertainties are dominated by the
PM uncertainties, with distance uncertainties making only a minimal
contribution.

In the present paper we use the observed PMs to determine both the
direction and size of the M31 velocity vector with respect to the MW,
and we use this knowledge with the Local Group timing argument (Kahn
\& Woltjer 1959; vdMG08; and Li \& White 2008) to estimate the LG
mass. We then compare the velocity and mass results to independent
estimates of the same quantities. For example, vdMG08 estimated the
transverse motion of M31 based on the kinematics of satellite galaxies
of M31 and the Local Group. Furthermore, the mass of the Local Group
has been estimated independently by adding up the individual masses of
the MW and M31, as estimated from various dynamical tracers
(e.g. Klypin \etal 2002; Watkins \etal 2010). By statistically
combining all the results we are able to build an improved and
comprehensive understanding of the dynamics and mass of the LG.

The outline of this paper is as follows. In Section 2 we use $N$-body
models of M31 and its prominent tidal substructures to calculate
predictions for the {\it internal} kinematics of M31 stars in the
three fields observed with HST. We use the results to correct the
transverse velocities measured with HST, to estimate the transverse
velocity of the M31 center-of-mass (COM). In Section 3 we revisit the
methods of vdMG08 to estimate the M31 transverse motion from the
kinematics of satellites, but with an expanded satellite sample. We
combine the results with the HST measurements to obtain a final
estimate for the M31 transverse motion. In Section 4 we derive the
corresponding space motion in the Galactocentric rest frame, taking
into account the latest insights about the solar motion in the MW. The
results are consistent with a radial orbit for M31 towards the MW. In
Section 5 we use the M31 motion to estimate the LG mass using the
timing argument. We find that the estimate is quite uncertain due to
cosmic scatter, and we show how a more accurate estimate can be
obtained by combining it with estimates of the individual MW and M31
masses. In Section 6 we consider the galaxy M33, the third most
massive galaxy of the Local Group (van den Bergh 2000), and we derive
its relative velocity with respect to M31. We also derive an estimate
for the mass of M33, and show that M33 is most likely bound to M31, as
is usually assumed. We use this knowledge to further refine our
estimate for the Local Group mass. In Section 7 we discuss and
summarize the main results of the paper. An Appendix presents a
discussion of various parameterizations used in the literature (and
the paper text) to quantify the dark halo density profiles and masses
of galaxies. Where necessary to compare the properties of Local Group
galaxies with predictions from cosmological simulations, we use a
Hubble constant $H_0 = 70 \kms \Mpc^{-1}$ and a matter density
$\Omega_m = 0.27$ (Jarosik \etal 2011).

This is the second paper in a series of three. Paper III (van der
Marel \etal 2012, in prep.) will present a study of the future orbital
evolution and merging of M31, M33, and the MW, using the velocities
and masses derived in the present paper as starting conditions.

\section{Correction for Internal Kinematics}
\label{s:intkin}

The PMs measured with HST in M31 fields contain contributions from
both the M31 COM motion, and from the internal kinematics of M31. In
each field, different fractions of the stars are contributed by
different structural components. Specifically, the galaxy has
different equilibrium components, including both a disk and spheroids
(bulge/halo). We will refer to these jointly as the ``base galaxy''.
The galaxy also contains material that is in the process of being
accreted. This includes in particular the material responsible for the
creation of the GSS (which in fact is spread out over a large fraction
of the projected area of the galaxy, and not just the actual position
of the Stream). To estimate the M31 COM motion, we need to know for
each field observed with HST both the fraction of the stars in each
component, and the transverse motion kinematics of those stars. The
fractional contributions can in principle be estimated purely
observationally from line-of-sight (LOS) velocity studies. However,
estimates of the transverse motion kinematics requires a full
dynamical model, since these motions are not directly constrained
observationally. We therefore resort to N-body models for M31 like
those previously constructed by some of us (e.g., Geehan \etal 2006;
Fardal \etal 2006, 2007, 2008) to understand various observed features
of M31.

\subsection{N-body models of M31 Structure}
\label{ss:Nbody}

The M31 model we use here is constructed from two separate but related
parts. The base galaxy is an N-body realization of a model of
the mass and light in M31 itself. The GSS component is a snapshot from
a dynamical N-body simulation of the formation of the GSS, performed
using the same mass model of M31. Taken together, these two components
reproduce reasonably well the features in M31 that are expected to
contribute to our HST fields.

The base galaxy, which is a slightly altered version of the model from
Geehan \etal (2006), contains bulge, disk, and halo components. The
bulge and disk are assumed to be mostly baryonic and therefore trace
the light. To the dark halo present in Geehan \etal (2006) we add a
stellar halo, which is necessary to reproduce the extended power-law
component that has been discovered in the halo regions (Guhathakurta
\etal 2005; Irwin \etal 2005; Kalirai \etal 2006b; Chapman \etal 2006;
Ibata \etal 2007). We assume this stellar halo follows the mass
distribution of the dark halo, although it contains only a tiny
fraction of that mass. When added together, these components satisfy
the surface-brightness profiles of M31's bulge, disk, and halo regions
reasonably well. Most importantly for this study, they also satisfy a
series of kinematic constraints, including the disk rotation curve,
the bulge velocity dispersion, and constraints on the halo mass from
statistical tracers such as globular clusters, planetary nebulae,
satellite galaxies, and red giant stars. We created the particle
realization of this model using the ZENO library (Barnes 2011).

The GSS component is created by simulating the disruption of a
satellite galaxy, in a fixed potential corresponding to the mass model
just discussed. The model is an updated version of that found in
Fardal \etal (2007), to which we refer the reader for a physical
discussion. This model uses a spherical progenitor, although it is
possible that the progenitor may in fact have been a disk galaxy
(Fardal \etal 2008). After starting at large radius with carefully
chosen initial conditions, the satellite disrupts at its first
pericentric passage. The model is evolved using the PKDGRAV tree
N-body code (Stadel 2001) for nearly 1~Gyr, until it forms orbital
wraps closely resembling features in M31 including the GSS itself, and
the NE and W shelves. We refer to all the particles generated by this
component as the GSS component, regardless of where they are on the
sky.

All the parameters of the base model, N-body simulations, and
data-model comparisons are presented in Fardal \etal (2012, in prep.).
However, many properties and details are similar to preceding papers
(Geehan \etal 2006; Fardal \etal 2006, 2007, 2008). Besides the
morphological evidence, the GSS model satisfies a set of observational
constraints, including the detailed kinematic pattern in the W Shelf,
the precise sky position of the GSS, the distance to various fields
along the GSS (McConnachie \etal 2003), and their peak LOS velocities
(Ibata \etal 2004; Guhathakurta \etal 2006; Kalirai \etal 2006a;
Gilbert \etal 2009). We do not use the observed color-magnitude
diagrams (CMDs) of the HST PM fields to constrain the models, since
those are not easily decomposed into distinct structural
components. In fact, the CMDs of the HST spheroid field and the HST
stream field are strikingly similar, given that they are believed to
be dominated by different structural components (Brown \etal 2006).

\begin{figure}
\epsscale{0.6}
\plotone{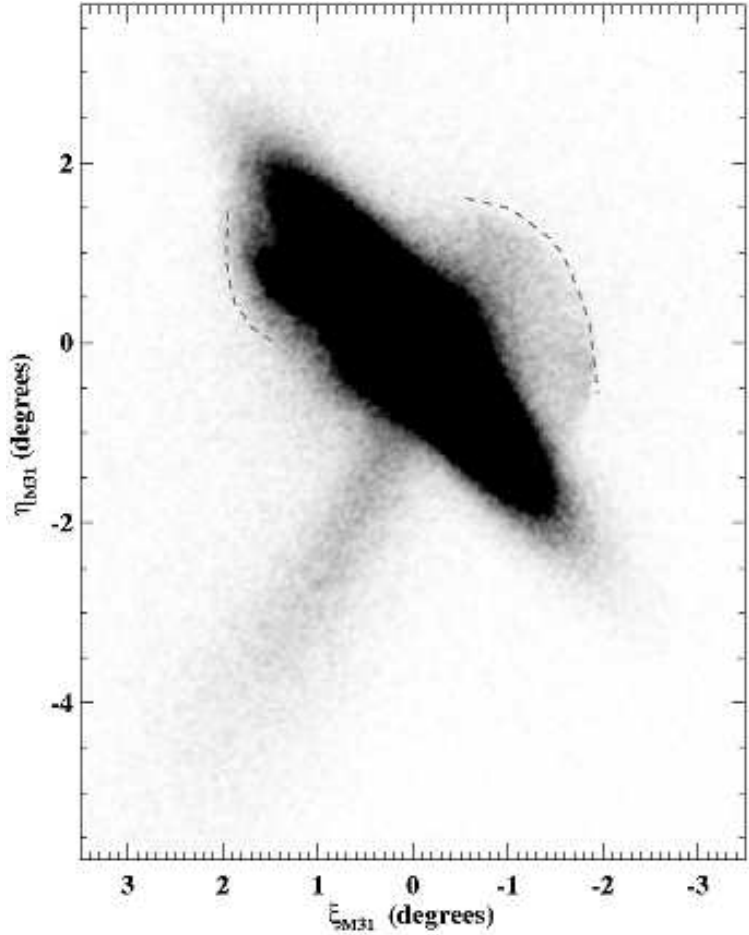}
\figcaption{Smoothed projected view of the N-body model used to
  calculate the internal M31 PM kinematics for our HST fields from
  Paper~I. A standard sky projection is used, with North up and East
  to the left. The GSS is visible South-East of the galaxy center, and
  the observed positions of the North-East and Western shelf are shown
  with dashed outlines. This image can be compared to star count maps
  of giant stars in M31, such as that reproduced in Figure 1 of
  Paper~I, which show very similar features.
\label{f:Nbodyproj}}
\end{figure}

\begin{figure}
\epsscale{0.6}
\plotone{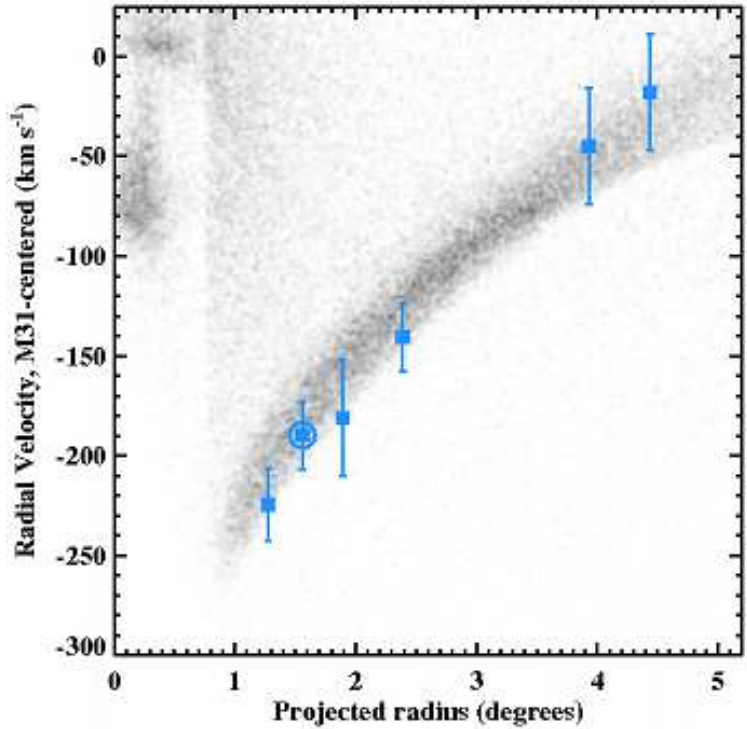}
\figcaption{Smoothed view in LOS velocity vs.~projected distance space
of the N-body model used to calculate the internal M31 PM kinematics
for our HST fields from Paper~I. Only particles with $Y <
-0.75^{\circ}$ are shown (located South-East of the galaxy center),
where $Y$ is a cartesian coordinate along the projected galaxy minor
axis. The dark band in the figure is due to the GSS, while the base
galaxy contributes most of the remaining particles. The GSS location
matches the observed peak LOS velocity of the GSS as a function of
radius (blue points; Ibata \etal 2004; Guhathakurta \etal 2006;
Kalirai \etal 2006a; Gilbert \etal 2009), including that measured in
the HST stream PM field of Paper~I (circle).
\label{f:Nbodyvel}}
\end{figure}

Figure~\ref{f:Nbodyproj} shows a smoothed projected view of the N-body
model. The GSS is visible South-East of the galaxy center, and the
North-East and Western shelf are emphasized with dashed outlines.
This image can be compared to star-count maps of giant stars in M31,
which show the same features (e.g., Ibata \etal 2005; Gilbert \etal
2009; or Paper I, which also shows the location of our PM
fields). Figure~\ref{f:Nbodyvel} shows a smoothed view of the N-body
model in LOS velocity vs.~projected distance space, for particles
South-East of the galaxy center. There is good agreement between the
outline of the GSS in this representation (dark band in the figure),
and the observed peak LOS velocity of the GSS as a function of radius
(blue points), including that measured in the HST stream PM field of
Paper~I (circle).

\subsection{Proper-Motion Corrections for HST Fields}
\label{ss:PMcor}

The predictions of the N-body model for M31 are summarized in
Table~\ref{t:intkin} for each of the three HST fields. The quantity
$f_{\rm base}$ is the fraction of the stars that belongs to the base
galaxy, and $f_{\rm GSS}$ is the fraction of the stars that belongs to
the GSS.  The average velocities in the LOS, W, and N directions are
listed for both the base and the GSS components, and also for their
properly weighted average (``all''). The quantities were extracted
over fields that are somewhat larger than the HST fields (up to 10
arcmin from the field center), to decrease the N-body shot noise. All
velocities are expressed in a reference frame in which M31 is at rest.

The average internal transverse velocity kinematics of the M31 stars
in the HST fields $(v_W,v_N)(all)$ are generally small, always below
125 km/s in absolute value. There are several reasons for this. In the
HST spheroid field we are sampling primarily the spheroidal components
of M31. At large radii these have limited mean rotation (Dorman \etal
2012). In the HST disk field we are primarily sampling the M31 disk,
which has a sizeable circular velocity ($\sim 250$ km/s).  However,
M31 has a large inclination. So along the major axis, most of the
rotation is seen along the LOS, and not in the transverse
direction. In the HST stream field we are primarily sampling the GSS,
which has a significant mean three-dimensional velocity ($254$
km/s). However, the inclination of the stream is such that most of the
velocity is seen along the LOS. Moreover, some 20\% of the stars in
the stream field do not belong to the GSS, but mostly to the spheroid
component.

For all three fields, the contribution to the observed transverse
motion from the internal kinematics of M31 is similar to or smaller
than the random errors in the HST measurements. Even significant
fractional changes in the model predictions therefore do not strongly
affect our final results. It therefore did not seem worthwhile in the
context of the present study to further refine the model.
Nonetheless, it is worth pointing out that the model is far from
perfect, and that there are some salient features of M31 that it fails
to reproduce. For example, the model with a spherical progenitor
overestimates the contribution of GSS particles on their first orbital
wrap to fields along the minor axis (especially those more distant
than our spheroid field). Also, LOS velocity studies of the GSS have
revealed a secondary cold component (in addition to the GSS and base
galaxy; e.g., Kalirai \etal 2006a; Gilbert \etal 2007, 2009) which is
not reproduced by our model. This could be, e.g., from a severely
warped disk component, or from wrapped-around material of a GSS loop
not included in our model. And finally, some authors have proposed
models for the structure of M31's outer and accreted components that
differ from those in our models (e.g., Ibata \etal 2005; Gilbert \etal
2007).

To estimate the transverse M31 COM motion from the data for each HST
field, we first subtract the contribution from internal M31 kinematics
($v(all)$) from the measurement ($v(HST)$). We then correct for the
effect of viewing perspective as described in van der Marel \etal
(2002) and vdM08. This corrects for the fact that at the position of
each field, the M31 COM systemic LOS and transverse velocity,
respectively, are not exactly aligned along the {\it local} LOS and
transverse directions. The corrections are small (below 10 km/s in
absolute value), because all fields are located within 2 degrees of
the M31 center. The final estimates are listed in Table~\ref{t:intkin}
as $(v_W,v_N)(COM)$, and are summarized also at the top of
Table~\ref{t:Andvel}. In propagating the uncertainties, we assigned an
uncertainty of 20 km/s per coordinate to the model of the M31 internal
kinematics. This number need not be known accurately, since the final
uncertainties are always dominated by measurement errors in the PMs.

The results for the three different fields are mutually consistent
with each other at the 1$\sigma$ level (see also
Paper~I\footnote{Paper~I defines a $\chi^2$ quantity, $\chi^2_3$, that
assesses the extent to which measurements for different fields agree.
Table~\ref{t:Andvel} implies $\chi^2_3 = 3.5$, for $N_{\rm DF} = 4$
degrees of freedom.}). This justifies the use of a straightforward
weighted average to combine the results, which gives $(v_W,v_N)(COM) =
(-162.8 \pm 47.0 , -117.2 \pm 45.0)$ km/s. For comparison, the direct
weighted average of the HST observations, with no corrections for
internal M31 kinematics, is $(v_W,v_N)(HST) = (-154.1 \pm 44.9 ,
-112.9 \pm 42.9)$ km/s. Clearly, the corrections for internal
kinematics make only a small difference for the final transverse
velocity estimate. The fact that the differences are below 10 km/s is
due to our combination of results for well-chosen fields, since the
per-field corrections are much larger than this.

\section{Transverse Velocity from Satellite Kinematics}
\label{s:satkin}

In vdMG08 we presented several methods for estimating the space motion
of M31 from the kinematics of satellites, which assume that the
satellites of M31 and the LG on average follow their motion through
space. The M31 transverse velocity derived in that paper has random
error bars of 34 to 41 km/s. This is somewhat smaller than what we
have obtained here from the HST PM measurements, although the
systematic error bars on the vdMG08 values may be larger (because the
underlying methodology makes more assumptions). Either way, these
results remain of considerable interest as an independent constraint
on the M31 space motion. We therefore update here the results from
vdMG08 using additional satellite data that has become available more
recently.

\subsection{Constraints from Line-of-Sight Velocities of M31 Satellites}
\label{ss:los}

The first method of vdM08 is based on the fact that any transverse
motion of M31 induces an apparent solid body rotation in the
line-of-sight velocity field of its satellites, superposed on their
otherwise primarily random motions. The amplitude and major axis of
the rotation field are determined by the size and direction of M31's
transverse motion. In vdM08 we constrained the M31 transverse motion
by fitting the velocities of 17 M31 satellites with known
line-of-sight velocities.

For the present study we added the satellites listed in
Table~\ref{t:satlos}. These are objects that previously either did not
have LOS velocity measurements available, or which had not yet been
discovered. This includes six dSph galaxies: And XI, XIII, XV, XVI,
XXI, and XXII. Three other recently discovered dSph galaxies, And
XVII, XIX, XX, have not yet had their LOS velocities measured. As in
vdMG08, we leave out And XII and XIV, because their large negative LOS
velocities with respect to M31 indicate that they may be falling into
M31 for the first time (Chapman \etal 2007; Majewski \etal 2007). We
also leave out the more recently discovered And XVIII, which may be
too distant from M31 to be directly associated with it (McConnachie
\etal 2008). We do include And XXII, even though it may be a satellite
of M33 rather than M31 (Martin \etal 2009; Tollerud \etal 2012).  We
note that And IV is not included in our combined sample because it is
a background galaxy (Ferguson \etal 2000), while And VIII is not a
galaxy at all (Merrett \etal 2006). For all dSphs, including those
from Paper~I and not listed in Table~\ref{t:satlos}, we used the newly
measured LOS velocities from Tollerud \etal (2012), where
available. Otherwise, the values listed in Paper~I or the sources
listed in Table~\ref{t:satlos} were used. Our new sample in
Table~\ref{t:satlos} also includes the 8 globular clusters of M31 that
lie at projected distances $> 40$ kpc and have known LOS velocities.

We repeated the vdMG08 analysis, using the combined sample of the 17
satellites in Table~1 of vdMG08 and the 14 satellites in
Table~\ref{t:satlos}. The implied space motion of M31 is listed in
Table~\ref{t:Andvel} in the row labeled ``M31 satellites''. The result
for $(v_W,v_N)$ differs from that derived in vdMG08 by $(-40.1,13.4)$
km/s. This is considerably smaller than the error bars in the result
of $(144.1,85.4)$ km/s. The addition of the 14 new satellites has not
decreased the error bars on the result. This is in part because most
satellites are observed relatively close to M31, so that any
solid-body rotation signal would be small. As before, the new result
is roughly consistent at the 1$\sigma$ level with zero transverse
motion. So no solid-body rotation component is confidently detected,
which in turn implies that M31 cannot have a very large transverse
motion. The fits imply a one-dimensional velocity dispersion for the
satellite sample of $\sigma_{\rm sat} = 84.8 \pm 10.6$ km/s. This is
$8.5 \kms$ larger than the value derived in vdMG08, which again is
within the uncertainties.

% modelfit2_out/modelfit2.10out1

\subsection{Constraints from Proper Motions of M31 Satellites}
\label{ss:pm}

The second method of vdM08 is based on the M31 satellites M33 and IC
10. These galaxies have accurately known PMs from water-maser
observations (Brunthaler \etal 2005, 2007). The three-dimensional
velocity vectors of these galaxies give an estimate of the M31 space
motion to within an accuracy of $\sigma_{\rm sat}$ per
coordinate. Transplanting the M33 and IC 10 velocity vectors to the
position of M31, followed by projection onto the local LOS, W, and N
directions, yields the results listed in Table~\ref{t:Andvel} in the
rows labeled ``M33 PM'' and ``IC 10 PM''. These are identical to what
was derived in vdMG08, but with slightly larger uncertainties (due to
the increased estimate of $\sigma_{\rm sat}$).

\subsection{Constraints from Line-of-Sight Velocities of Outer Local Group 
Galaxies}
\label{ss:outerlos}

The third method of vdMG08 is based on the line-of-sight velocities of
Local Group satellites that are not individually bound to the MW or
M31. In vdMG08 the method was applied to 5 satellites (see their Table
2). The Cetus dSph (RA$=6.54597^{\circ}$, DEC$=-11.04432^{\circ}$) was
excluded because of lack of knowledge of its LOS velocity at the
time. For the present study we have rerun the analysis including
Cetus, using $v_{\rm LOS} = -87 \pm 2$ km/s from Lewis \etal
(2007). Its distance $D = 755 \pm 23 \kpc$ (McConnachie \etal 2005)
places Cetus at $D_{\rm bary} \approx 600 \kpc$ from the Local group
barycenter. With addition of the Cetus dSph to the vdMG08 analysis,
the implied space motion of M31 is listed in Table~\ref{t:Andvel} in
the row labeled ``Outer LG Galaxies''. The result for $(v_W,v_N)$
differs from that derived in vdMG08 by $(-14.9,-13.5)$ km/s. This is
considerably smaller than the error bars in the result of
$(58.0,52.5)$ km/s.

\subsection{Comparison and Combination of Constraints}
\label{ss:combine}

In general, modeling of satellite galaxy dynamics can be complicated
for a variety of reasons, especially when the goal is to estimate
galaxy masses: the satellite system may not be virialized, with
continueing orbital evolution (Mateo \etal 2008) or infall (e.g.,
Chapman \etal 2007; Majewski \etal 2007); the distribution of
satellite orbits may not be isotropic (e.g., Watkins \etal 2010);
satellites on large-period orbits are not expected to be randomly
distributed in orbital phase (e.g., Zaritsky \& White 1994);
satellites may have correlated kinematics (e.g., van den Bergh 1998);
and the three-dimensional distribution of satellites may not be
spherical (Koch \& Grebel 2006) or symmetric (McConnachie \& Irwin
2006). However, many of these potential issues do not affect the
analysis that we have presented here and in vdMG08 to estimate the M31
transverse velocity. Sections~\ref{ss:los} and~\ref{ss:pm} only assume
that the M31 satellites are drawn from a distribution that has the
same mean velocity as M31, and which has no mean rotation.
Section~\ref{ss:outerlos} only assumes that the LG satellites are
drawn from a distribution that has the same mean velocity as the LG
barycenter. Virialized equilibrium, isotropy, random phases, or
symmetry are not required. Nonetheless, there is always the
possibility that residual systematics may have affected the
results. To get a handle on this, we have compared in detail the
results for the M31 transverse velocity from the different techniques.

The $(v_W,v_N)$ for M31 inferred from the different methods in
Sections~\ref{ss:los}--\ref{ss:outerlos} are in mutual agreement to
within the uncertainties. The same was true also in the original
analysis of vdMG08. Since the methods and the underlying data are
quite different for the various estimates, this in itself is a direct
indication that any residual systematics cannot be large. Since the
results from the different methods are in agreement, it is reasonable
to take their weighted average, as listed in Table~\ref{t:Andvel}.
The result for $(v_W,v_N)$ differs from that derived in vdMG08 by
$(-19.0,-7.1)$ km/s. This is considerably smaller than the error bars
in the result of $(40.7,36.6)$ km/s, so the new analysis presented
here has not significantly altered the results previously derived by
vdMG08.

\begin{figure}
\epsscale{0.6}
\plotone{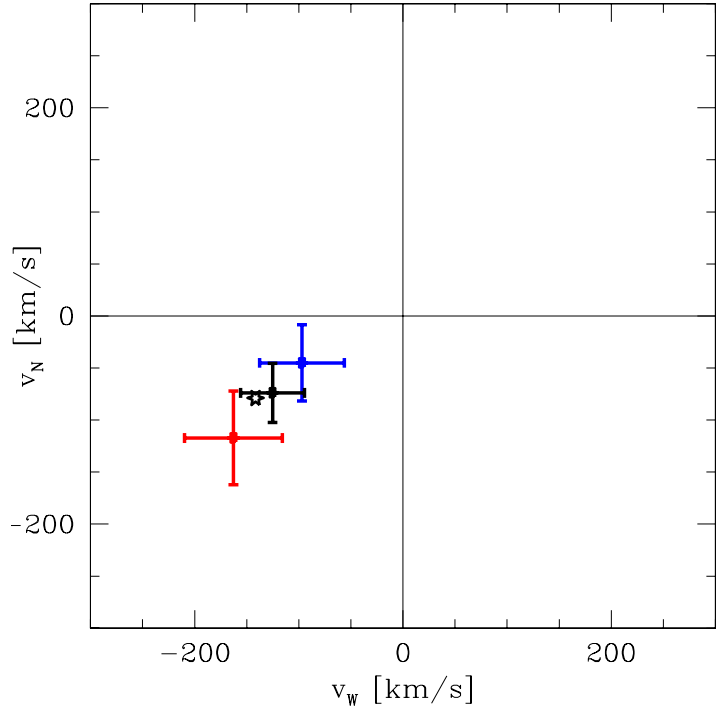}
\figcaption{Estimates of the M31 heliocentric transverse velocity in
  the West and North directions. Data points with error bars are from
  Table~\ref{t:Andvel}. Red: Weighted average of HST proper-motion
  measurements, corrected for internal kinematics
  (Section~\ref{s:intkin}). Blue: Weighted average of methods based on
  satellite kinematics (update of vdMG08 result;
  Section~\ref{s:satkin}). Black: Overall weighted average of all
  measurements. The starred symbol indicates the transverse velocity
  that corresponds to a radial orbit for M31 with respect to the Milky
  Way. The measurements are consistent with a radial orbit.
\label{f:vwvn}}
\end{figure}

An even stronger check on any residual systematics is provided by
Figure~\ref{f:vwvn}. It compares the weighted average of the HST PM
measurements (with corrections for internal kinematics) from
Section~\ref{s:intkin} (as listed in Table~\ref{t:Andvel}) with the
weighted average from the updated vdMG08 analysis. The difference
between these results is $(\Delta v_W, \Delta v_N) = (-65.8 \pm 62.2 ,
-72.1 \pm 58.0)$.  This means that the results are consistent within
the uncertainties: the probability of a residual this large occurring
by chance in a {\it two-dimensional} Gaussian distribution is
26\%. Since the methods employed are totally different, and have quite
different scopes for possible systematic errors, this is very
successful agreement. This suggests not only that there are no large
residual systematics in the results from the satellite kinematics, but
also that there are no large residual systematics in the M31 PM
analysis. This is a very important cross-check, since the
displacements on which our PM measurements are based are below $0.01$
detector pixels (see Paper~I for a detailed discussion of the
systematic error control in the PM analysis).

Since the HST PM analysis and the satellite kinematics analysis yield
statistically consistent results for the M31 transverse velocity, it
is reasonable to take the weighted average of the results from the two
methodologies. This yields 
\begin{equation}
  (v_W,v_N) = (-125.2 \pm 30.8 , -73.8 \pm 28.4) \kms ,
\end{equation}
as listed in the bottom row of Table~\ref{t:Andvel} and shown in black
in Figure~\ref{f:vwvn}.  This is the final result that we use for the
remainder of our analysis.

\section{Space Motion}
\label{s:spacevel}

\subsection{Galactocentric Rest Frame and Solar Motion}
\label{ss:solar}

As in vdMG08, we adopt a Cartesian coordinate system $(X,Y,Z)$, with
the origin at the Galactic Center, the $Z$-axis pointing towards the
Galactic North Pole, the $X$-axis pointing in the direction from the
Sun to the Galactic Center, and the $Y$-axis pointing in the direction
of the Sun's Galactic rotation. We choose the origin of the frame to
be at rest (the Galactocentric rest frame), and we wish to calculate
the velocity of galaxies in this frame. This requires knowledge of the
solar velocity in the Milky Way, since the solar reflex motion
contributes to any {\it observed} velocities (such as the {\it
heliocentric} velocities listed in Table~\ref{t:Andvel}).

In vdMG08 we adopted the standard IAU values (Kerr \& Lynden-Bell
1986) for the distance of the Sun from the Galactic Center $R_0 = 8.5
\kpc$, and the circular velocity of the Local Standard of Rest (LSR),
$V_0 = 220$ km/s. Neither of these quantities has historically been
known particularly accurately though, and their exact values continue
to be debated. Recently, a number of new methodologies have become
available. These provide new insights into $R_0$ and $V_0$, and we
therefore use the results from these studies here.

Some of the best constraints on $R_0$ now come from studies of the
orbits of stars around the Sgr A* supermassive black hole. Gillessen
\etal (2009) obtained $R_0 = 8.33 \pm 0.35 \kpc$ (consistent also with
Ghez \etal 2008). Most of the available constraints on the velocity
$V_0$ are actually constraints on the ratio $V_0/R_0$. The best
constraint on this ratio now comes from the observed PM of Sgr A*,
since the black hole is believed to be at rest in the galaxy to within
$\sim 1$ km/s. Reid \& Brunthaler (2004) obtained that $(V_0 + V_{\rm
  pec}) / R_0 = 30.2 \pm 0.2 \kms \kpc^{-1}$. Here $V_{\rm pec}$ is
the peculiar velocity of the Sun in the rotation direction. In vdMG08
we adopted the solar peculiar velocity from Dehnen \& Binney
(1998). However, there is now increasing evidence that $V_{\rm pec}$
from that study (and other studies) is too small by $\sim 7$ km/s. We
adopt here the more recent estimates from Sch\"onrich, Binney, \&
Dehnen (2010): $(U_{\rm pec}, V_{\rm pec}, W_{\rm pec}) = (11.1,
12.24, 7.25)$, with uncertainties of $(1.23,2.05,0.62) \kms$ (being
the quadrature sum of the random and systematic errors). Combination
of these results implies that $V_0 = 239.3 \pm 10.3$ km/s,
significantly larger than the canonical IAU value of 220 km/s. The
uncertainty is dominated entirely by the uncertainty in $R_0$, and the
errors in $V_0$ and $R_0$ are highly correlated.

Observations of masers in high-mass star-formation regions in the MW
have been used to argue independently for a value of $V_0$ in excess
of the canonical $220 \kms$ (Reid \etal 2009). However, McMillan \&
Binney (2010) showed that these data by themselves do not strongly
constrain the Galactic parameters. On the other hand, McMillan (2011)
showed that when combined with the other constraints described above
through detailed models, the maser data do help to constrain $R_0$
more tightly, and therefore $V_0$. He obtained: $R_0 = 8.29 \pm 0.16
\kpc$ and $V_0 = 239 \pm 5$ km/s. These are the values we adopt here.

\subsection{M31 Space Motion}
\label{ss:M31space}

% spacemotion_out/spacemotion.6out1 and spacemotion.7out1

Based on the adopted M31 distance and solar parameters, the position
of M31 in the Galactocentric rest frame is 
\begin{equation}
  {\vec r}_{\rm M31} = (-378.9,612.7,-283.1) \kpc .
\end{equation}
The velocity of the Sun projects to $(v_{\rm sys}, v_W, v_N)_{\odot} =
(191.9, 142.5, 78.5) \kms$ at the position of M31. Since one observes
the reflex of this, these values must be {\it added} to the observed
M31 velocities to obtain its velocity in the Galactocentric rest
frame. The velocity vector corresponding to the observed COM LOS
velocity $v_{\rm LOS} = -301 \pm 1 \kms$ (vdMG08) and the final
weighted average $(v_W,v_N)$ given in Table~\ref{t:Andvel},
transformed to the Galactocentric rest frame, is then 
\begin{equation}
  {\vec v}_{\rm M31} = (66.1 \pm 26.7 , -76.3 \pm 19.0 , 45.1 \pm 26.5) \kms .
\end{equation}
The errors (which are correlated between the different components)
were obtained by propagation of the errors in the individual position
and velocity quantities (including those for the Sun) using a
Monte-Carlo scheme.

If the transverse velocity of M31 in the Galactocentric rest frame,
$V_{\rm tan}$, equals zero, then M31 moves straight towards the Milky
Way on a purely radial (head-on collision) orbit. This orbit has
$(v_W,v_N)_{\rm rad} = (-141.5 \pm 3.0, -78.8 \pm 1.7) \kms$ (this is
approximately the reflex of the velocity of the Sun quoted above,
because the lines from the Sun to M31 and from the Galactic Center to
M31 are almost parallel). The listed uncertainty is due to propagation
of the uncertainties in the solar velocity vector. The radial orbit is
indicated as a starred symbol in Figure~\ref{f:vwvn}. The velocity
${\vec v}_{\rm M31}$ calculated in the previous paragraph corresponds
to a total velocity
\begin{equation}
  |{\vec V}_{\rm M31}| = 110.6 \pm 7.8 \kms . 
\end{equation}
The radial velocity component is
\begin{equation}
   V_{\rm rad,M31} = -109.2 \pm 4.4 \kms ,
\end{equation}
and the tangential velocity component is
\begin{equation}
   V_{\rm tan,M31} = 17.0 \kms \qquad 
   (1\sigma\ {\rm confidence\ region:\ } V_{\rm tan,M31} \leq 34.3 \kms).
\end{equation}
The uncertainties were calculated as in vdMG08, using a flat Bayesian
prior probability for $V_{\rm tan}$. These results imply that the
velocity of M31 is statistically consistent with a radial orbit at the
1$\sigma$ level.\footnote{If the IAU value $V_0 = 220$ km/s is used
for the LSR circular velocity and the Dehnen \& Binney (1998) values
are used for the solar peculiar velocity (as in vdMG08), then a radial
orbit has $(v_W,v_N)_{\rm rad} = (-126.6, -71.4) \kms$. With these
assumptions, the inferred velocity of M31 is still statistically
consistent with a radial orbit at the 1$\sigma$ level.}

It has been known for a long time that the transverse velocity of M31
is probably $V_{\rm tan, M31} \lesssim 200 \kms$. The large scale
structure outside the LG does not provide enough tidal torque to have
induced a much larger transverse motion, subsequent to the radial
expansion started by the Big Bang (Gott \& Thuan 1978; Raychaudhury \&
Lynden-Bell 1989). Li \& White (2008) used the $\Lambda$CDM
cosmological Millenium simulation to identify some thousand galaxy
pairs that resemble the MW-M31 pair in terms of morphology, isolation,
circular velocities, and radial approach velocity. The median
tangential velocity of the pairs was $V_{\rm tan} = 86 \kms$, with
$24$\% of the pairs having $V_{\rm tan} < 50 \kms$ (see their
figure~6). Therefore, the observed M31 tangential velocity is somewhat
below average compared to cosmological expectation, but it is not
unusually low. The exact reason why the tangential velocity of the
MW-M31 pair has ended up below-average is not clear, but it may be
related to the details of the growth history and the local environment
of the LG.

Peebles \etal (2001) showed that at values $V_{\rm tan, M31} \lesssim
200 \kms$, many velocities can be consistent with the observed
positions and velocities of galaxies in the nearby Universe. Our new
observational result that $V_{\rm tan,M31} \leq 34.3 \kms$ at
1$\sigma$ confidence therefore significantly reduces the parameter
space of possible orbits. Peebles \etal (2011) recently proposed a
model for the history and dynamics of the LG in which $V_{\rm tan,
M31} = 100.1 \kms$ and $(v_W,v_N) = (-240.5 , -63.1) \kms$. This is
inconsistent with our final velocity estimates at $>3\sigma$
confidence.

\section{Local Group Mass}
\label{s:mass}

The velocity vector of M31 with respect to the Milky Way constrains
the mass of the Local Group through the so-called ``timing argument''
(Kahn \& Woltjer 1959; Lynden-Bell 1981, 1999; Einasto \& Lynden-Bell
1982; Sandage 1986; Raychaudhury \& Lynden-Bell 1989; Kroeker \&
Carlberg 1991; Kochanek 1996). Recent applications of this method were
presented in vdMG08 and Li \& White (2008). In Section~\ref{ss:timing}
we provide a revised estimate of the Local Group timing mass using the
new insights into the M31 velocity vector from
Section~\ref{ss:M31space}. In Section~\ref{ss:posterior} we combine
the result in statistical fashion with results from other independent
methods for constraining the Local Group mass.

Different studies often quote different mass quantities. However, for
proper use and comparison it is important to transform all
measurements to a common definition. For each galaxy, the total mass
is dominated by a very extended dark halo. Common characterizations of
dark matter halos are summarized in Appendix~\ref{s:darkprof}. The
density profile is often modeled as an NFW profile (Navarro \etal
1997; eq.~[\ref{rhoNFW}]) or a Hernquist (1990; eq.~[\ref{rhoHern}])
profile. The former has infinite mass, while the latter has finite
mass $M_H$. Common characterizations of halo masses also include the
virial mass $M_{\rm vir}$ enclosed within radius $r_{\rm vir}$
(eq.~[\ref{Rvir}]), the mass $M_{\rm 200}$ enclosed within the radius
$r_{\rm 200}$ (eq.~[\ref{qdef}]), or the mass $M(r)$ enclosed within
some given physical radius $r$ in kpc
(eqs.~[\ref{massNFW},\ref{massHern}]).

In our discussion of galaxy masses, we transform all results into
$M_{\rm vir}$ estimates. The transformation requires knowledge of the
density profile, which for this purpose we assume to be of the NFW
form with known concentration $c_{\rm vir}$ (eq.~[\ref{massNFW}]). For
the MW and M31 we take $c_{\rm vir} = 10 \pm 2$, based on a
combination of specific models (Klypin \etal 2002; Besla \etal 2007),
and cosmological simulation results (Neto \etal 2007; Klypin \etal
2011).  This implies $M_{\rm 200}/M_{\rm vir} = 0.839 \pm 0.014$
(eq.~[\ref{M200}]).

\subsection{Timing Argument}
\label{ss:timing}

Under the assumption of Keplerian motion, the relative orbit of M31
and the MW is determined by four parameters: the total mass $M_{\rm
tot}$; the semi-major axis length $a$ (or alternatively, the orbital
period $T$); the orbital eccentricity $e$; and the current position
within the orbit, determined by the eccentric anomaly $\eta$. In turn,
four observables are available to constrain the orbit: the current M31
distance $D$; the radial and tangential M31 velocities in the
Galactocentric rest frame, $V_{\rm rad,M31}$ and $V_{\rm tan,M31}$;
and the time $t$ since the last pericenter passage, which should be
equal to the age of the Universe $t = 13.75 \pm 0.11 \Gyr$ (Jarosik
\etal 2011; when the matter of both galaxies originated together in
the Big Bang). There are as many observables as unknowns, so the
orbital parameters and $M_{\rm LG}$ can be determined uniquely. The
implied value of $M_{\rm tot}$ is called the ``timing mass''. Most
commonly, the relevant equations are solved under the assumption of a
radial orbit ($e=1$ and $V_{\rm tan,M31} = 0$), but this assumption is
not necessary when a measurement of $V_{\rm tan,M31}$ is actually
available (e.g., vdMG08).

For the M31 space motion derived above, the ``timing mass'' $M_{\rm
tot,timing} = (4.27 \pm 0.53) \times 10^{12} \Msun$. When a radial
orbit is assumed (which is consistent with the data) then $M_{\rm
tot,timing} = (4.23 \pm 0.45) \times 10^{12} \Msun$ (this is somewhat
smaller, because any transverse motion increases the timing mass). The
listed uncertainties are the RMS scatter of Monte-Carlo simulations as
in vdMG08, which take into account all the observational
uncertainties. The black curve in the top panel of
Figure~\ref{f:masses} shows the complete probability histogram for the
radial orbit case.

These timing mass results are about $1.0$--$1.3 \times 10^{12} \Msun$
lower than what was obtained by, e.g., vdMG08 and Li \& White
(2008). This can be viewed as an improvement, since previous estimates
of $M_{\rm tot}$ appeared anomalously high compared to independent
estimates of the masses of the individual M31 and MW galaxies (as
summarized in vdMG08). Some of the decrease in timing mass is due to
the fact that $V_{\rm tan,M31}$ found here is slightly smaller than in
vdMG08. However, most of the decrease in timing mass is not due to the
new HST measurements, but due to the new values for the solar motion
used here.  The solar velocity in the $Y$-direction, $v_Y = V_0 +
V_{\rm pec}$, is $251.2$ km/s in our calculations here. By contrast,
it was $26$ km/s lower in the calculations of vdMG08. The $Y$
component of the solar motion projects predominantly along the LOS
direction towards M31, and not the W and N directions. The component
of the solar motion in the LOS direction is therefore $20.2$ km/s
higher than what it was in vdMG08.  As a consequence, in the
Galactocentric rest frame, M31 approaches the MW with a radial
velocity that is $20.2$ km/s slower than what is was in vdMG08. This
slower approach implies a lower timing mass.

The timing argument equations are based on a simple Keplerian
formalism. To assess how accurate this argument is in a cosmological
context one must make comparisons to $N$-body simulations (Kroeker \&
Carlberg 1991). Li \& White (2008) did this for the currently favored
$\Lambda$CDM cosmology, using results from the Millennium simulation.
They identified simulated galaxy pairs like the MW-M31 system, with
known masses, and quantified the accuracy of the {\it radial orbit}
timing argument. They found that the timing argument has very little
bias, when viewed as an estimate of the sum $M_{\rm tot, 200}$ of the
galaxy's $M_{\rm 200}$ values, but significant scatter. They
quantified this ``cosmic scatter'', by averaging over pairs with all
possible transverse velocities. However, not surprisingly, their
Figure~6 shows that the scatter increase with $V_{\rm tan}$. Since we
now know that M31 actually has a low value of $V_{\rm tan}$, it is
more appropriate to quantify the cosmic scatter by restricting the
statistics to pairs in the simulation with low $V_{\rm tan}$. We
measured by hand from their Figure~6 all pairs with $V_{\rm tan} \leq
50$ km/s, and extracted the ratio $M_{\rm tot, 200}/M_{\rm tot,
timing}$. We folded the probability distribution of these ratios into
our Monte-Carlo scheme for estimating the total mass from the radial
orbit timing argument. This yields the estimate $M_{\rm tot, 200} =
(4.14 \pm 1.36) \times 10^{12} \Msun$, which has a three times larger
uncertainty than what is implied by observational errors alone. This
can be converted into an estimate for the summed virial masses using
to formulae of Appendix~A, which yields
\begin{equation}
  M_{\rm tot, vir} = (4.93 \pm 1.63) \times 10^{12} \Msun \qquad 
  {\rm (timing\ argument)} .
\end{equation}
This is our final estimate from the timing argument, which takes into
account observational uncertainties, cosmic bias, and cosmic
scatter. The red curve in the top panel of Figure~\ref{f:masses} shows
the complete probability histogram for $M_{\rm tot, vir}$.

% HSTPM2/masses/combine.2outlog

\subsection{Combination with Other Milky Way and M31 Mass Constraints}
\label{ss:posterior}

The best alternative method for estimating the mass of the Local Group
is to add up estimates of the masses of the individual M31 and MW
galaxies.\footnote{The other method of estimating $M_{\rm tot}$ from
the size of the Local Group turn-around radius yields estimates that
tend to be biased low. A radial infall model is generally assumed,
which is almost certainly an oversimplification (see vdMG08).}
Estimates for the masses of these galaxies were already summarized in
vdMG08, so here we highlight primarily some more recent results.

\begin{figure}
\epsscale{1.0}
\plotone{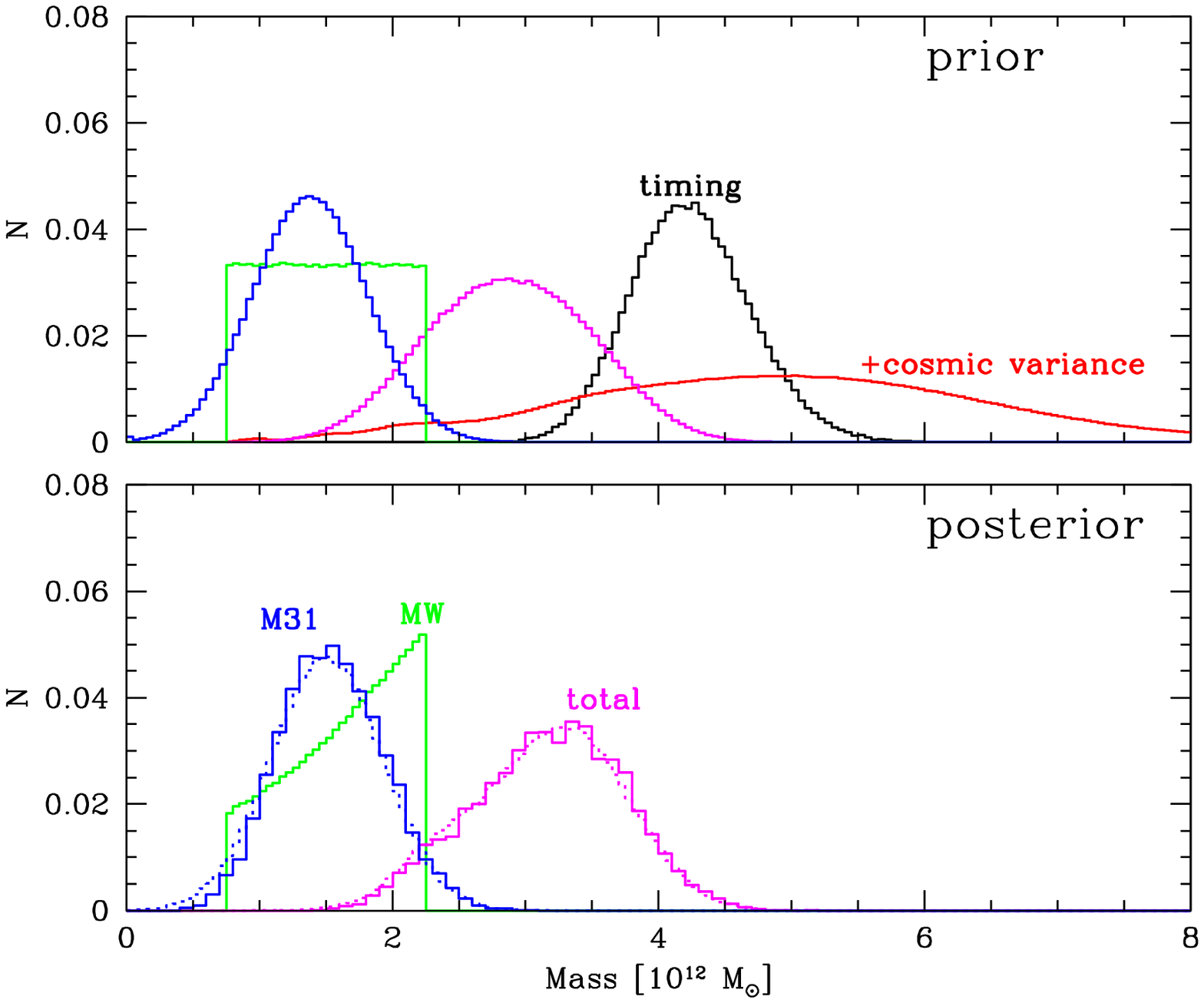}
\figcaption{Probability distributions for the mass of the Milky Way,
M31, and their sum $M_{\rm tot}$. The top panel shows prior
probability distributions based on several lines of evidence.  M31
(blue) and MW (green): probability distributions based on studies of
these galaxies as discussed in the text; Sum of M31 and MW (magenta);
Timing argument $M_{\rm tot}$ with inclusion of observational errors
(black), and with additional inclusion of cosmic variance from Li \&
White (2008; red). The bottom panel shows with the same color coding
posterior probability distributions, obtained by combining constraints
as described in the text. Dotted: knowledge of the individual galaxies
combined with the the timing argument; the timing argument does not
help much to constrain the masses, due to its large cosmic variance.
Solid: Requiring also that M33 and M31 be a bound pair; this reduces
the probability of low M31 masses.
\label{f:masses}}
\end{figure}

Watkins \etal (2010) studied the kinematics of M31 satellites and
found that the mass within $300 \kpc$ is determined fairly robustly,
$M_{\rm M31}(300 \kpc) = (1.40 \pm 0.43) \times 10^{12} \Msun$. The
quoted uncertainty is the quadrature sum of the random error of $0.40
\times 10^{12} \Msun$, and a systematic uncertainty of $0.15 \times
10^{12} \Msun$ due to the assumed velocity anisotropy of the
satellites. There may be other systematic uncertainties in the
analysis, but these are more difficult to quantify and are neglected
here. At this mass and with the relevant halo concentrations, $M (300
\kpc) /M_{\rm vir} = 1.018 \pm 0.002$ (see
Appendix~\ref{s:darkprof}). Hence, $M_{\rm M31,vir} = (1.38 \pm 0.43)
\times 10^{12} \Msun$. We use this to set a Gaussian probability
distribution for $M_{\rm M31, vir}$ in our discussion below (blue
curve in top panel of Figure~\ref{f:masses}). This is consistent with
the study of Klypin \etal (2002), which folded in a wider range of
observational constraints, and obtained successful models with dark
halo masses $M_{\rm M31,dark,vir}$ of either $1.43 \times 10^{12}
\Msun$ or either $1.60 \times 10^{12} \Msun$. This corresponds to a
total mass of $M_{\rm M31,vir} = 1.52 \times 10^{12} \Msun$ or $1.69
\times 10^{12} \Msun$, respectively, after adding in also the combined
stellar mass of the M31 disk and bulge. The Watkins \etal results are
also consistent with the recent results of Tollerud \etal (2012). They
applied a mass estimator calibrated on cosmological simulations to the
M31 satellite kinematics and obtained $M_{\rm M31,vir} =
1.2_{-0.7}^{+0.9} \times 10^{12} \Msun$.

Watkins \etal (2010) showed that mass estimates for the mass of the MW
from satellite kinematics are much more uncertain. This is due to the
unknown velocity anisotropy, combined with the fact that we see most
satellites almost radially from near the Galactic Center. Good mass
estimates therefore need to fold in a more diverse set of
observational constraints. Moreover, uncertainties are reduced
significantly by assuming that the radial profile of the dark matter
is known, and follows a cosmologically motivated parameterization.
McMillan (2011) used such methods to obtain a dark halo mass $M_{\rm
MW, dark, 200} = (1.26 \pm 0.24) \times 10^{12} \Msun$. This
corresponds to $M_{\rm MW, dark, vir} = (1.50 \pm 0.29) \times 10^{12}
\Msun$. This is consistent with the study of Klypin \etal (2002), who
favored $M_{\rm MW, dark, vir} = 1.0 \times 10^{12} \Msun$, but showed
that reasonable models with $M_{\rm MW, dark, vir} = 2.0 \times
10^{12} \Msun$ can be constructed as well. Adding the combined stellar
mass of the MW disk and bulge to obtain the total $M_{\rm MW, vir}$
adds $\sim 0.06 \times 10^{12}$ to these values. The rapid motion of
the Magellanic Clouds and Leo I have been used to argue for masses at
the high end of this range of values (e.g., Zaritsky \etal 1989;
Shattow \& Loeb 2008; Li \& White 2008; Boylan-Kolchin \etal 2011).
However, the underlying assumptions in these arguments cause
significant uncertainties. Based on the range of results in the
literature, we adopt here, fairly arbitrarily, a flat probability
distribution for $M_{\rm MW,vir}$ between $0.75$ and $2.25 \times
10^{12}\Msun$ (green curve in top panel of
Figure~\ref{f:masses}). This distribution has the same mean ($1.50
\times 10^{12} \Msun$) as inferred by McMillan (2011), and the same
dispersion ($0.43 \times 10^{12} \Msun$) as we use for M31, but with a
broader, flatter shape.\footnote{There do exist models in the
literature that yield or use higher mass estimates for M31 and the MW
than we use here. This includes estimates based on halo occupation
distributions (e.g., Guo \etal 2010) or the timing argument (e.g.,
Loeb \etal 2005; Cox \& Loeb 2008). It should be kept in mind though
that such estimates are statistical in nature.  Cosmic scatter must
therefore be taken into account, and this yields large uncertainties
(e.g., Li \& White 2008; Guo \etal 2010). It is therefore important
that any mass estimate {\it for an individual galaxy}, as opposed to
an ensemble of galaxies, also take into account the actually observed
resolved properties, rotation curves, and satellite kinematics.}

We use the listed probability distributions for the individual M31 and
MW masses as priors; this also sets a prior probability distribution
for $M_{\rm tot,vir} \equiv M_{\rm M31,vir} + M_{\rm MW,vir}$ (magenta
curve in top panel of Figure~\ref{f:masses}). We then fold in the
timing argument results to determine posterior probability
distributions, as follows. We draw a random mass from the probability
distribution for $M_{\rm tot,vir}$ derived from the timing argument
(red curve in top panel of Figure~\ref{f:masses}). We then draw a
random $M_{\rm MW,vir}$ from its prior distribution. We then calculate
the corresponding $M_{\rm M31,vir} = M_{\rm tot,vir} - M_{\rm
MW,vir}$, and its probability $p$ given the prior distribution for
$M_{\rm M31,vir}$. This set of values is then accepted or rejected in
Monte-Carlo sense, depending on the probability $p$. We thus build up
posterior probability distributions for $M_{\rm MW,vir}$, $M_{\rm
M31,vir}$, and $M_{\rm tot,vir}$, which are shown as dotted lines with
the same colors in the bottom panel of Figure~\ref{f:masses}.

The posterior distribution for $M_{\rm M31,vir}$ is still roughly
Gaussian, but its average has increased from $(1.38 \pm 0.43) \times
10^{12} \Msun$ to $(1.51 \pm 0.42) \times 10^{12} \Msun$. The
posterior distribution for $M_{\rm MW,vir}$ is not flat like its
prior, but skewed towards higher masses. Its average has increased
from $1.50$ to $1.63 \times 10^{12} \Msun$. The likelihood of MW and
M31 masses at the low end of the prior distributions is significantly
reduced in the posterior distributions. This is relevant, since some
previously reported mass estimates do fall on this low-mass end (e.g.,
Evans \etal 2000; Ibata \etal 2004). At the best-estimate virial
masses for the MW and M31, the corresponding virial radii are $308
\kpc$ and $300 \kpc$, respectively (eq.~[\ref{Rvir}]). Since the
distance between the galaxies is $D = 770 \pm 40 \kpc$, the virial
spheres are not currently overlapping.

The prior distribution for $M_{\rm tot,vir}$ corresponds to $(2.88 \pm
0.61) \times 10^{12} \Msun$, while its posterior distribution
corresponds to $(3.14 \pm 0.58) \times 10^{12} \Msun$. Therefore,
inclusion of the timing argument increases the estimate of the LG mass
by only $\sim 9$\%, due to the large cosmic variance. Since this is
considerably smaller than the prior uncertainties on $M_{\rm MW,vir}$
and $M_{\rm M31,vir}$, the timing argument does not in fact help much
to constrain the total LG mass, beyond what we already know from the
MW and M31 individually. We have found this to be a robust conclusion,
independent of the exact probability distributions adopted for M31 and
the MW, and independent of the exact solar and M31 motion adopted in
the timing argument.

% HSTPM2/masses/combine.2outlog

\section{Mass Constraints from M33}
\label{sec:M33}

The galaxy M33 is the most massive companion of M31 (e.g., van den
Bergh 2000). In the past decade, evidence has been found from both HI
(Braun \& Thilker 2004) and star-count maps (McConnachie \etal 2009)
for tidal features indicative of past interactions between these
galaxies. Models for these features such as those presented by
McConnachie \etal (2009) require that M33 be bound to
M31.\footnote{While the MW also has a massive companion, namely the
Large Magellanic Cloud, it is unclear whether the galaxies in this
pair form a bound system (Besla \etal 2007).} The galaxy M33 is one of
the few galaxies in the Local Group for which an accurate PM
measurement is available from VLBA observations of water
masers. Hence, combined with our new M31 results, the relative motion
of M33 with respect to M31 is now known with reasonable accuracy
(Section~\ref{ss:M33space}). The mass of M33 can also be estimated
independently (Section~\ref{ss:M33mass}). With knowledge of the
relative velocity and mass, the assumption that M33 is bound to M31
can be used to further refine our understanding of the M31 mass, and
hence the Local Group mass (Section~\ref{ss:bound}).
 
\subsection{M33 Space Motion}
\label{ss:M33space}

To establish the binding energy of the M31-M33 system, we need to know
the current position ${\vec r}_{\rm M33}$ and velocity ${\vec v}_{\rm
M33}$ of M33 in the Galactocentric rest frame. These were determined
in similar fashion as for M31 (see Section~\ref{ss:M31space}), but now
based on the following observables: a distance $D_{\rm M33} = 794 \pm
23 \kpc$ (McConnachie \etal 2004), line-of-sight velocity $v_{\rm
los,M33} = -180 \pm 1 \kms$ (vdMG08), and PM from water
masers as measured by Brunthaler \etal (2005) and discussed in
vdMG08. This yields ${\vec r}_{\rm M33} = (-476.1, 491.1, -412.9)
\kpc$, and ${\vec v}_{\rm M33} = (43.1 \pm 21.3 , 101.3 \pm 23.5 ,
138.8 \pm 28.1) \kms$. The observational errors in the Galactocentric
velocity of M33 are similar to those for M31 reported in
Section~\ref{ss:M31space}.

The positions of the three galaxies MW, M31 and M33 define a plane in
the Galactocentric rest frame. For simplicity, we will refer to this
plane as the ``trigalaxy plane''. It is of interest for understanding
the orbital evolution of the MW-M31-M33 system, to know how the M33
velocity vector is oriented with respect to this plane. To assess
this, we introduce a new Cartesian coordinate system $(X',Y',Z')$
based on the following definitions: the frame has the same origin as
the $(X,Y,Z)$ system (i.e., the Galactic Center); the $X'$-axis points
from the origin to M31 at $t=0$; the $Y'$-axis is perpendicular to the
$X'$-axis, and points from M31 to M33 as seen in projection from the
Galactic Center; and the $Z'$-axis is perpendicular to the $X'$- and
$Y'$-axes in a righthanded sense. With these definitions, the
trigalaxy plane is the $(X',Y')$ plane. Hence, let us refer to the
$(X',Y',Z')$ system as the ``trigalaxy coordinate system''.

Based on the position vectors ${\vec r}_{\rm M31}$ and ${\vec r}_{\rm
M33}$ from Sections~\ref{ss:M31space} and~\ref{ss:M33space}, the unit
vectors of the $(X',Y',Z')$ system can be expressed in $(X,Y,Z)$
coordinates as
\begin{eqnarray}
  {\vec u}_{X'} &= (-0.48958 ,  0.79153 , -0.36577) , \nonumber \\
  {\vec u}_{Y'} &= (-0.47945 , -0.60013 , -0.64029) , \nonumber \\
  {\vec u}_{Z'} &= (-0.72632 , -0.13810 ,  0.67331) .
\label{triunits}
\end{eqnarray}
If ${\vec r}$ is a vector expressed in Galactocentric $(X,Y,Z)$
coordinates, then the corresponding vector ${\vec r}'$ expressed in
the trigalaxy coordinate system is 
\begin{equation}
  {\vec r}' = (X',Y',Z') = ({\vec r} \cdot {\vec u}_{X'} ,
                            {\vec r} \cdot {\vec u}_{Y'} ,
                            {\vec r} \cdot {\vec u}_{Z'} ) , 
\label{tricoords}
\end{equation}
where $\cdot$ denotes the vector inner product. We use
equations~(\ref{triunits},\ref{tricoords}) as the fixed definition of
the $(X',Y',Z')$ system throughout this paper, even when we vary the
positions of the Sun, M31, and M33 within their observational
uncertainties.

Observational uncertainties of $\sim 20$--$30 \kms$ aside, the
velocities of M31 and M33 in trigalaxy coordinates are ${\vec v}'_{\rm
M31} = (-109.2, -15.5, -7.1) \kms$ and ${\vec v}'_{\rm M33} = (8.3,
-170.3, 48.2) \kms$. These vectors make angles with the $(X',Y')$
plane of only $-3.7^{\circ}$ and $15.8^{\circ}$, respectively.  By
definition, the MW galaxy currently has zero velocity in the
Galactocentric rest frame. However, the gravitational attraction from
M31 and M33 will set it in motion with a velocity directed in the
$(X',Y')$ plane. Hence, all three galaxies start out in the $(X',Y')$
plane, with velocity vectors that are close to this plane. This
implies that the orbital evolution of the entire MW-M31-M33 system
will happen close to the trigalaxy plane, with the ``vertical''
$Z'$-component playing only a secondary role. Detailed calculations of
the future orbital evolution and merging of the MW-M31-M33 system are
the topic of Paper~III.

% HSTPM2/coordinates1.out2

\subsection{M33 Mass}
\label{ss:M33mass}

The mass of M33 is not negligible with respect to that of M31. It is
therefore necessary to know the mass of M33 to determine whether the
M31-M33 system is bound. Corbelli (2003) modeled the rotation curve
and mass content of M33.  The rotation curve rises to $\sim 130 \kms$
out to the last data point at $15 \kpc$. Since the data do not reveal
a turnover in the rotation curve, both the halo concentration and
virial mass are poorly constrained (Fig.~6b of Corbelli
2003). Moreover, the rotation field is complex with significant
twisting (Corbelli \& Schneider 1997).  This complicates
interpretation in terms of circular motion. To estimate the M33 virial
mass it is therefore necessary to use more indirect arguments. For
this, we compare M33 to M31.

Corbelli (2003) used her rotation curve fits to estimate the
mass-to-light ratio of the M33 disk. From this, she inferred a stellar
mass 2.8--5.6 $\times 10^9 \Msun$ at $3\sigma$
confidence.\footnote{Corbelli's mass scale for $H_0 = 65 \kms
\Mpc^{-1}$ was transformed to the Hubble constant used here. The small
mass contribution from the nuclear component of M33 is well within the
quoted uncertainties. M33 has no bulge.} Higher values correspond to a
maximum-disk fit, while lower values correspond to a sub-maximal
disk. Guo \etal (2010) instead used the observed $B-V$ color of M33
with stellar population model predictions to estimate the
mass-to-light ratio. With an assigned uncertainty of 0.1 dex for this
method, one obtains $(2.84 \pm 0.73) \times 10^9 \Msun$. We
combine these methods into a single rough estimate $M_{\rm M33,*} =
(3.2 \pm 0.4) \times 10^9 \Msun$.

For M31, Klypin \etal (2002) used rotation-curve fits to estimate both
the disk and the bulge mass. The two models they present cover the
ranges $M_{\rm M31, disk} = 7.0$--$9.0 \times 10^{10} \Msun$ and
$M_{\rm M31, bulge} = 1.9$--$2.4 \times 10^{10} \Msun$. Upon
subtraction of the gas mass of $\sim 0.6 \times 10^{10}$ (van den
Bergh 2000), this yields $M_{\rm M31,*} = (8.3$--$10.8) \times 10^{10}
\Msun$. The Guo \etal (2010) method based on the galaxy $B-V$ color
yields instead $M_{\rm M31,*} = (7.0 \pm 1.8) \times 10^{10}
\Msun$. We combine these methods into a single rough estimate $M_{\rm
M31,*} = (7.9 \pm 0.9) \times 10^{10} \Msun$.

These estimates imply that $M_{\rm M33,*} / M_{\rm M31,*} = 0.041 \pm
0.007$.  This can be compared to the {\it baryonic} mass ratio implied
by the Tully Fisher relation, $(V_{\rm M33}/V_{\rm M31})^4$ (McGaugh
2005). With $V_{\rm M33} \approx 130 \kms$ (Corbelli \& Salucci 2000)
and $V_{\rm M31} \approx 250 \kms$ (Corbelli \etal 2010) this yields
$0.073$. This is consistent with the estimate of $M_{\rm M33,*} /
M_{\rm M31,*}$, if one takes in to account that in M33 the stars make
up only $\sim 57$\% of the baryonic mass, the rest being mostly in
neutral and molecular gas (Corbelli 2003).

Models of the halo occupation distribution of galaxies predict a
relation for $M_*/M_{\rm 200}$ as function of halo mass $M_{\rm 200}$,
when matching observed galaxy properties from the Sloan Digitial Sky
Survey to the properties of dark matter halos seen in simulations
(e.g., Wang \etal 2006; Guo \etal 2010). From
Section~\ref{ss:posterior} we have $M_{\rm M31,vir} = (1.50 \pm 0.38)
\times 10^{12} \Msun$, which corresponds to $M_{\rm M31,200} = (1.26
\pm 0.32) \times 10^{12} \Msun$. Combined with knowledge of the
observed $M_{\rm M33,*} / M_{\rm M31,*}$, this can be used to estimate
$M_{\rm 200}$ for M33, and hence the virial mass.\footnote{In relating
$M_{\rm 200}$ to $M_{\rm vir}$ for M33, we assume that $c_{\rm vir} =
10 \pm 2$, as we did for the MW and M33. While the lower mass of M33
would in principle lead one to expect a higher concentration, this is
not supported by fits to the rotation curve (Corbelli 2003).} This
yields $M_{\rm M33,vir} = (0.170 \pm 0.059) \times 10^{12} \Msun$
based on the Guo \etal relations, and $M_{\rm M33,vir} = (0.127 \pm
0.055) \times 10^{12} \Msun$ based on the Wang \etal relations. The
uncertainties were estimated using a simple Monte-Carlo scheme that
includes, in addition to the observational errors, the Gaussian cosmic
scatter of $\sim 0.2$ dex in stellar mass at fixed halo mass (Guo
\etal 2010).\footnote{This exceeds the observational errors in $M_{*}$
for both M33 and M31. As a result, it is not necessary for the present
method to have particularly robust estimates of these observational
uncertainties.} The difference in normalization between the
predictions from Wang \etal and Guo \etal is not well understood. So
we treat this as an additional model uncertainty, and allow all values
bracketed between the two relations with equal probability. This
yields as our final estimate $M_{\rm M33,vir} = (0.148 \pm 0.058)
\times 10^{12} \Msun$.\footnote{We could instead have used the mass
$M_{\rm M33,*}$ directly to estimate $M_{\rm M33,vir}$, with no
reference to M31. This yields $M_{\rm M33,vir} = (0.225 \pm 0.055)
\times 10^{12} \Msun$ based on the Guo \etal relations, and $M_{\rm
M33,vir} = (0.123 \pm 0.034) \times 10^{12} \Msun$ based on the Wang
\etal relations. Both of these estimates are consistent with what we
use here. However, there is a significant difference in absolute mass
normalization between the theoretical relations. The relations agree
better in a relative sense, which is why we prefer the method used
here. The latter uses only relative theoretical predictions, combined
with the kinematically determined virial mass for M31.}

Strictly speaking, the mass inferred from halo occupation
distributions is the so-called infall mass. Thus we assume that mass
loss to M31 has not yet been significant. On the other hand, the
uncertainty on $M_{\rm M33,vir}$ is significant. Also, our $M_{\rm
M33,vir}$ estimate falls below what is implied by direct application
of the Guo \etal relations. Therefore, significant mass loss would not
be inconsistent with the range of masses we explore here.

\subsection{Mass Implications of a Bound M31-M33 Pair}
\label{ss:bound}

To assess the likelihood, given the data, that M31 and M33 are bound,
we set up mass and velocity combinations in Monte-Carlo sense. The
initial masses $M_{\rm vir}$ for both galaxies were drawn as in
Sections~\ref{ss:posterior} and~\ref{ss:M33mass}. The initial
phase-space coordinates were drawn as in Sections~\ref{s:spacevel}
and~\ref{ss:M33space}. This scheme propagates all observational
distance and velocity uncertainties and their correlations, including
those for the Sun\footnote{Uncertainties in the RA and DEC of M31 and
  M33 are negligible and were ignored.}. For each set of initial
conditions we calculated the binding energy of the M33-M31 system. The
M33-M31 system was found to be bound in 95.3\% of cases. Therefore,
our observational knowledge of the masses, velocities, and distances
of these two galaxies indicates that indeed, they most likely form a
bound pair.

% HSTPM2/orbit_out/orbit78.log

The observation of tidal features associated with M33 independently
implies that M31 and M33 are likely a bound pair (Braun \& Thilker
2004; McConnachie \etal 2009). If we enforce this as a prior
assumption, then this affects our posterior estimates of the M31 and
M33 masses. To enforce this assumption, we merely need to remove from
our Monte-Carlo scheme those initial conditions in which M33 and M31
are not bound. Figure~\ref{f:masses}b shows as solid histograms the
posterior distributions after application of this additional prior.
The main effect is to disallow some of the initial conditions in which
$M_{\rm M31}$ (blue curve) is on the low end of its probability
distribution. The average and RMS mass increase from $M_{\rm M31, vir}
= (1.51 \pm 0.42) \times 10^{12} \Msun$ to $(1.54 \pm 0.39) \times
10^{12} \Msun$. The mass distribution of M33 is not appreciably
affected. The posterior mass distribution for $M_{\rm tot,vir} = M_{\rm
MW, vir} + M_{\rm M31, vir}$ is shown as the cyan histogram. Its
average and RMS are 
\begin{equation}
   M_{\rm tot,vir} = (3.17 \pm 0.56) \times 10^{12} \Msun \qquad
   {\rm (final\ estimate)} .
\end{equation}
This is similar to the result from Section~\ref{ss:posterior}, which
was $M_{\rm tot,vir} = (3.14 \pm 0.58) \times 10^{12} \Msun$. Hence,
the assumption that M33 must be bound to M31 does not help much to
reduce the uncertainties in the LG mass, since the fact that they are
bound is already implied at high confidence by the observed
velocities.

% HSTPM2/orbit_out/analyze78.com

The probability distributions of M31 and M33 distances and velocities
are not appreciably affected by the additional prior that M31 and M33
be bound. The average positions and velocities in the Galactocentric
rest frame remain the same to within $1 \kpc$ and a few km/s,
respectively, after the unbound orbits are removed.

\section{Discussion and Conclusions}
\label{s:conc}

We have presented the most accurate estimate to date of the transverse
motion of M31 with respect to the Sun. This estimate was made possible
by the first PM measurements for M31, made using HST, and presented in
Paper~I. We have combined these measurements with other insights to
constrain the transverse motion of M31 with respect to the MW. We have
used the resulting motion to improve our understanding of the mass of
the Local Group, and its dominant galaxies M31 and the MW.

The HST PM measurements from Paper~I pertain to three fields in M31.
The PM for each field contains contributions from three components:
the M31 COM motion, the known viewing perspective, and the internal
kinematics of M31. To correct for the contributions from internal
kinematics, we have constructed detailed $N$-body models. The models
include both the equilibrium disk, bulge, spheroid, and dark-halo
components, as well as the material from a tidally disrupted satellite
galaxy that is responsible for the GSS. Even though the stars in M31
move at velocities of hundreds of km/s, the internal-kinematics
corrections to the observed PMs averaged over all fields is quite
small ($\lesssim 25 \kms$, well below the random uncertainties in the
measurements). This is largely due to the known properties of the
carefully chosen field locations, and the galaxy components that they
sample.

The resulting M31 transverse motion should be largely free from
systematic errors, based on the many internal consistency checks built
into our PM program, as discussed in Paper I. This includes the fact
that the observations for the three different fields, including
observations with different instruments at different times, all yield
statistically consistent estimates for the M31 COM motion.
Nonetheless, an entirely independent check on the results is obtained
by comparison to the M31 transverse motion estimates implied by the
methods from vdMG08, which are based exclusively on the kinematics of
the {\it satellite galaxies} of M31 and Local Group.

Instead of using the published results from vdMG08 directly, we have
redone their analysis using expanded satellite samples, including new
data that has become available in recent years. The end result is
similar to what was already published by vdMG08. More importantly, the
result is statistically consistent with that obtained from the HST PM
program. Since the methods employed are totally different, and have
very different scopes for possible systematic errors, this is very
successful agreement.  This gives added confidence in both results,
and also suggests that a further reduction in the uncertainties can be
obtained by taking the weighted average of both methods. This yields
$(v_W,v_N) = (-125.2 \pm 30.8 , -73.8 \pm 28.4)$, which is our final
estimate for the {\it heliocentric} transverse motion of M31. The
uncertainties in this result are similar to what has been obtained
from VLBA observations of water masers in the M31 satellites M33 and
IC10 (Brunthaler \etal 2005, 2007).

To understand the motion of M31 with respect to the MW, it is
necessary to correct for the reflex motion of the Sun. We adopted the
most recent insights into the solar motion within the Milky Way. These
imply an azimuthal motion for the Sun (the sum of the LSR motion and
the solar peculiar velocity) of $\sim 250$ km/s, which is $\sim 25$
km/s higher than what has typically been used in previous
studies. This implies a radial approach velocity of M31 with respect
to the Milky Way of $V_{\rm rad,M31} = -109.2 \pm 4.4 \kms$, which is
$\sim 20$ km/s slower than what has typically been used in previous
studies. The best estimate for the tangential velocity component is
$V_{\rm tan,M31} = 17.0 \kms$, with 1$\sigma$ confidence region
$V_{\rm tan,M31} \leq 34.3 \kms$. Hence, the velocity of M31 is
statistically consistent with a radial (head-on collision) orbit
towards the MW at the 1$\sigma$ level.

The new insights into the motion of M31 with respect to the MW allowed
us to revise estimates of the Local Group timing mass, as presented
most recently by vdMG08 and Li \& White (2008). This yields $M_{\rm
tot, timing} = 4.23 \times 10^{12} \Msun$ for an assumed radial orbit,
with a random error from observational uncertainties of $0.45 \times
10^{12} \Msun$.  This result is $\sim 20$\% lower than typically found
in previous studies, due to the lower $V_{\rm rad,M31}$ used here. We
calibrated the timing mass as in Li \& White (2008) based on
cosmological simulations. However, we selected from their galaxy pairs
in the Millennium simulation only those with low $V_{\rm tan,M31}$,
for consistency with the observations. This yields $M_{\rm tot,vir}
\equiv M_{\rm MW, vir} + M_{\rm M31, vir} = (4.93 \pm 1.63) \times
10^{12} \Msun$, where the uncertainty now includes cosmic scatter
(which dominates over random errors).

We have presented a Bayesian statistical analysis to combine the
timing mass estimate for $M_{\rm tot, vir}$ with estimates for the
individual masses of M31 and the MW obtained from other dynamical
methods. For the individual masses we used relatively broad priors
that encompass most values suggested in the literature. Even then, the
cosmic scatter in the timing mass is too large to help much in
constraining the mass of the Local Group. Its main impact is to
increase by $\sim 10$\% the mass estimates already known for the
individual galaxies (and their sum).

In an attempt to further refine the M31 and Local Group mass
estimates, we have studied the galaxy M33. Its known PM allowed us to
study the relative motion between M31 and M33. A range of arguments
suggests that the mass of M33 is $\sim 10$\% of the M31 mass. The
masses and relative motions of M31 and M33 indicate that they are a
bound pair at $95$\% confidence. Observational evidence for tidal
deformation between M33 and M31 suggests that the small 5\%
probability for unbound pairs, as allowed by the observational
uncertainties, may not be physical. This makes low values for the M31
mass unlikely, and hence increases the expectation value for the Local
Group mass, but only by $\sim 1$\%. Our final estimate for the Local
Group mass from all considerations is $M_{\rm tot,vir} = (3.17 \pm 0.57)
\times 10^{12} \Msun$.

The velocity vectors between M31, M33 and the MW are all closely
aligned with the plane that contains these galaxies. Paper III
presents a study of the future orbital evolution and merging of these
galaxies, using the velocities and masses derived here as starting
conditions.

\acknowledgements

Support for Hubble Space Telescope proposal GO-11684 was provided by
NASA through a grant from STScI, which is operated by AURA, Inc.,
under NASA contract NAS 5-26555. The PKDGRAV code used in
Section~\ref{ss:Nbody} was kindly made available by Joachim Stadel and
Tom Quinn, while the ZENO code used in that section was kindly made
available by Josh Barnes. The authors are grateful to T.~J.~Cox for
contributing to the other papers in this series, and to the anonymous
referee for useful comments and suggestions.

\appendix

\section{Dark Halo Profiles, Masses, and Sizes}
\label{s:darkprof}

Spherical infall models show that a virialized mass $M_{\rm vir}$ has
an average overdensity $\Delta_{\rm vir}$ compared to the average
matter density of the Universe. The virial radius $r_{\rm vir}$
therefore satisfies ${\overline \rho}_{\rm vir} \equiv 3 M_{\rm vir} /
4 \pi r_{\rm vir}^3 = \Delta_{\rm vir} \Omega_m \rho_{\rm crit}$, or
in physical units (Besla \etal 2007)
\begin{equation}
  r_{\rm vir} = 206 h^{-1} \kpc 
                \left ( {{\Delta_{\rm vir} \Omega_m} \over {97.2}}
                      \right )^{-1/3}
                \left ( {{M_{\rm vir}} \over {10^{12} h^{-1} \Msun}}
                      \right )^{1/3} .
\label{Rvir}
\end{equation}
For the cosmological parameters used here, $h = 0.7$ and $\Omega_m =
0.27$, one has $\Delta_{\rm vir} = 360$ (Klypin \etal 2011). 

Dark halo density profiles in cosmological simulations are well
described by an NFW density profile (Navarro \etal 1997),
\begin{equation}
  \rho_N (r) = \rho_s x^{-1} (1+x)^{-2} , \qquad x \equiv r/r_s  .
\label{rhoNFW}
\end{equation}
The enclosed mass is 
\begin{equation}
  M_N (r) = 4 \pi \rho_s r_s^3 f(x) 
          = M_{\rm vir} f(x) / f(c_{\rm vir}) , \qquad
  f(x) = \ln(1+x) - { {x} \over {1+x} } ,
\label{massNFW}
\end{equation}
where the concentration is defined as $c_{\rm vir} = r_{\rm vir} /
r_s$. The average enclosed mass density equals
\begin{equation}
  {\overline \rho}_N (r) \equiv 3M_N(r) / 4 \pi r^3 = 
     3 \rho_s (r_s/r)^3 f(x) . 
\label{rhoavNFW}
\end{equation}

Another characteristic radius that is often used is the radius
$r_{200}$ so that the average enclosed density is 200 times the
critical density of the Universe, ${\overline \rho}_{\rm 200} \equiv 3
M_{\rm 200} / 4 \pi r_{\rm 200}^3 = 200 \rho_{\rm crit}$, where
$M_{\rm 200}$ is the enclosed mass. It follows from the respective
definitions that
\begin{equation}
  q \equiv {\overline \rho}_{\rm 200} / {\overline \rho}_{\rm vir} = 
  (200 / \Delta_{\rm vir}) \Omega_m^{-1} ,
\label{qdef}
\end{equation}
which yields $q = 2.058$ for the cosmological parameters used here. This 
exceeds unity, and therefore $r_{\rm 200} < r_{\rm vir}$ and 
$M_{\rm 200} < M_{\rm vir}$. For the NFW profile, $r_{\rm 200}$ is
the solution of the equation ${\overline \rho}_N (r_{\rm 200}) = 
q {\overline \rho}_N (r_{\rm vir})$, which implies  
\begin{equation}
  c_{\rm 200}/c_{\rm vir} = 
                  \left  ( { {f(c_{\rm 200})} \over {q f(c_{\rm vir})} }
                  \right )^{1/3}
\label{c200}
\end{equation}
where $c_{\rm 200} \equiv r_{\rm 200}/r_s$. This equation can be
quickly solved numerically using fixed point iteration, starting from
an initial guess for $c_{\rm 200}$ on the right hand side. The
corresponding mass ratio is 
\begin{equation}
  M_{\rm 200}/M_{\rm vir} = f(c_{\rm 200})/f(c_{\rm vir}) .
\label{M200}
\end{equation}

As discussed in Springel \etal (2005), it is often convenient for
numerical reasons to model dark halos with a Hernquist (1990) profile.
This is what we will do in our exploration of the orbital evolution of
the MW-M31-M33 system in Paper~III. In this case the density profile
is
\begin{equation}
  \rho_H (r) = \left ( { {M_H} \over {2 \pi a^3} } \right )
               y^{-1} (1+y)^{-3} , \qquad y \equiv r/a  .
\label{rhoHern}
\end{equation}
Here $M_H$ is the total mass of the system, which is finite, unlike for the 
NFW profile. The enclosed mass is 
\begin{equation}
  M_H (r) = M_H y^2 (1+y)^{-2} .
\label{massHern}
\end{equation}
The Hernquist profile has the same density as the NFW profile for $r
\rightarrow 0$ if 
\begin{equation}
  M_H = 2 \pi \rho_s a^2 r_s .
\end{equation}
We can choose the scale radius $a$ so that the enclosed mass 
of the NFW and Hernquist profiles is the same for some radius ${\tilde r}$,
$M_N({\tilde r}) = M_H({\tilde r})$. This implies 
\begin{equation}
  a/r_s = \left \lbrace
             [2 f({\tilde x})]^{-1/2} - (1/{\tilde x}) 
          \right \rbrace^{-1} ,
\label{aHern}
\end{equation}
where ${\tilde x} \equiv {\tilde r}/r_s$.
The corresponding total mass of the Hernquist profile satisfies
\begin{equation} 
  M_H / M_{\rm vir} = (a/r_s)^2 / [2 f(c_{\rm vir})]  
\label{Mhern}
\end{equation}
If we choose ${\tilde x} = c_{\rm 200}$, then the NFW and Hernquist
profiles have the same enclosed mass $M_{\rm 200}$ within $r_{\rm
200}$.\footnote{This is what Springel \etal (2005) aimed to
achieve. However, their equation~(2) is only an approximation to
equation~(\ref{aHern}), so they do not actually achieve this
equality.}. We denote by $a_{\rm 200}$ the corresponding value of 
$a$ from equation~(\ref{aHern}), and by $M_{H,{\rm 200}}$ the 
corresponding value of $M_H$ from equation~(\ref{Mhern}). If 
instead we choose ${\tilde x} = c_{\rm vir}$, 
then the NFW and Hernquist profiles have the same enclosed mass $M_{\rm vir}$ 
within $r_{\rm vir}$. In this case we denote by $a_{\rm vir}$ the 
corresponding value of $a$ from equation~(\ref{aHern}), and 
by$M_{H,{\rm vir}}$ the corresponding value of $M_H$ from 
equation~(\ref{Mhern}).

As an example, we consider a halo with $c_{\rm vir} = 10$. This yields
$c_{\rm 200} = 7.4$, $M_{\rm 200}/M_{\rm vir} = 0.84$, $a_{\rm
200}/r_s = 2.01$, $M_{H,{\rm 200}}/M_{\rm vir} = 1.36$, $a_{\rm
vir}/r_s = 2.09$, and $M_{H,{\rm vir}}/M_{\rm vir} = 1.46$.
  
%%%%%%%%%%%%%%
% Reference List
%%%%%%%%%%%%%%%

%%%%%%%%%%%%%%%
% TABLES
%%%%%%%%%%%%%%%

% Start tables on a new page
\clearpage

%%% TABLE 1 %%%

\begin{deluxetable}{llrrr}
\tabletypesize{\small}
\tablecaption{M31 Transverse Velocity: Proper Motions and Internal Kinematics
\label{t:intkin}} 
\tablehead{& & \colhead{\qquad Spheroid Field} & 
\colhead{\qquad Disk Field} & \colhead{\qquad Stream Field}}
\startdata
\multicolumn{5}{l}{{\it HST PM Measurements}}\\
\hline
$v_W$ (HST)   &  km/s & $-167.2 \pm 60.2$ & $-194.6 \pm 89.8$ & $-65.3 \pm 101.5$ \\
$v_N$ (HST)   &  km/s & $-137.2 \pm 56.2$ & $-38.0 \pm 89.1$ & $-130.3 \pm  99.3$ \\
% HSTPM/inplanerot4.out???7
\hline
\multicolumn{5}{l}{{\it M31 Internal Kinematics Model}}\\
\hline
$f$ (base)            &      & 0.738 & 0.922 &  0.195 \\
$f$ (GSS)             &      & 0.262 & 0.078 &  0.805 \\
$v_{\rm LOS}$ (base)  & km/s &  -1.1 & 219.8 &    4.8 \\
$v_{\rm LOS}$ (GSS)   & km/s &  71.9 & -82.3 & -185.0 \\
$v_{\rm LOS}$ (all)   & km/s &  18.0 & 196.3 & -148.0 \\
$v_W$ (base)          & km/s &  11.0 & -25.9 &   12.0 \\      
$v_W$ (GSS)           & km/s &  22.7 & -82.6 &   73.4 \\
$v_W$ (all)           & km/s &  14.1 & -30.3 &   61.4 \\
$v_N$ (base)          & km/s & -14.3 & -49.3 &   -7.1 \\        
$v_N$ (GSS)           & km/s &   2.5 &  41.8 &  157.5 \\
$v_N$ (all)           & km/s &  -9.9 & -42.2 &  125.4 \\
% HSTPM2/fardal4.txt
\hline
\multicolumn{5}{l}{{\it M31 COM Motion (HST PMs + Internal Kinematics Model + 
Viewing Perspective)}}\\
\hline
$v_W$ (COM)     &  km/s & $-179.1 \pm 64.1$ & $-158.0 \pm 92.4$ & $-126.3 \pm 103.6$ \\
$v_N$ (COM)     &  km/s & $-122.6 \pm 60.0$ & $  -0.5 \pm 91.3$ & $-247.5 \pm 102.1$ \\
% HSTPM/inplanerot4.out???7

\enddata
\tablecomments{Kinematical quantities for the
three HST fields observed in Paper~I. The top part of the table lists
the HST PM measurements, transformed to km/s using $D=770 \pm 40
\kpc$. The middle part gives the predictions of the M31 internal
kinematics model described in Section~\ref{s:intkin}. Predictions are
split into two components, the ``base'' equilibrium galaxy model
(disk, bulge, and halo), and the accreted ``GSS'' tidal stream
component. The fraction $f$ gives the amount contributed by each
component. Predictions for all model stars (independent of component),
are also listed (i.e., averages suitably weighted by the corresponding
fractions $f$). Average velocities are given in the line-of-sight
(LOS), West, and North directions. The directions are defined at the
center of each field. The internal kinematics velocities are 
expressed in a frame in which M31 is at rest. The bottom part of the
table gives the estimates for the center-of-mass (COM) motion of M31
that result from correcting the HST measurements for both 
internal kinematics and the viewing perspective. For the COM velocities,
the West and North directions are always defined at the COM.\looseness=-2}
\end{deluxetable}

%%% TABLE 2 %%%

\begin{deluxetable}{llrrrrl}
\tabletypesize{\small}
\tablecaption{Addition to vdMG08 M31 Satellite Galaxy Sample\label{t:satlos}} 
\tablehead{
\colhead{Name\qquad\qquad} & \colhead{Type\qquad\qquad} & \colhead{\qquad\qquad $\rho$} & 
\colhead{\qquad\qquad $\Phi$} & \colhead{\qquad\qquad $v_{\rm los}$} \\
 & & deg & deg & km/s \\
(1) & (2) & (3) & (4) & (5) }
\startdata
B517          & GC        &  3.29 &   77.48 & -272 $\pm$ 54 (G07) \\         
Mac-GC1       & GC        &  3.39 & -115.38 & -219 $\pm$ 15 (G07) \\         
B514          & GC        &  4.04 & -145.58 & -456 $\pm$ 23 (G05) \\         
EC4           & GC        &  4.39 &  135.88 & -288 $\pm$  2 (C09) \\   
B516          & GC        &  4.76 &   28.44 & -181 $\pm$  5 (G07) \\         
B518          & GC        &  5.74 & -110.08 & -200 $\pm$ 48 (G07) \\         
And XV        & dSph      &  6.84 &  114.86 & -323 $\pm$  1 (T12) \\         
B519          & GC        &  7.35 &  165.67 & -268 $\pm$ 47 (G07) \\         
And XI        & dSph      &  7.50 &  174.23 & -462 $\pm$  4 (T12) \\         
And XIII      & dSph      &  8.46 &  166.90 & -185 $\pm$  2 (T12) \\     
Mar-GC1       & GC        &  8.50 &  168.61 & -312 $\pm$ 17 (G07) \\         
And XXI       & dSph      &  9.00 &  -78.35 & -361 $\pm$  6 (T12) \\
And XVI       & dSph      &  9.50 &  158.04 & -367 $\pm$  3 (T12) \\         
And XXII      & dSph      & 16.06 &  141.60 & -127 $\pm$  3 (T12) \\ 
% M31satel.dat3j
% modelfit2_out/modelfit2.10out1
\enddata
\tablecomments{The sample of additional M31 satellites, which was
combined with the sample from Table~1 of vdMG08 for the modeling of
Section~\ref{ss:los}. Column~(1) lists the name of the object and
column~(2) its type. Objects labeled ``GC'' are distant globular
clusters. Columns~(3) and~(4) define the position on the sky: $\rho$
is the angular distance from M31 and $\Phi$ is the position angle with
respect to M31 measured from North over East, calculated from the sky
positions (RA,DEC) as in van der Marel \etal (2002). The satellites in
the table are sorted by their value of $\rho$. Column~(5) lists the
heliocentric line-of-sight velocity and its error. The source of the
measurement is listed in parentheses --- Collins \etal (2009): C09;
Tollerud \etal (2012): T12; Galleti \etal (2005): G05; Galleti \etal
(2007): G07. Sky positions were obtained from the listed sources or
the NASA Extragalactic Database.}
\end{deluxetable}

%%% TABLE 3 %%%

\begin{deluxetable}{rrrr}
\tabletypesize{\small}
\tablecaption{M31 Center-of-Mass Heliocentric Velocity Estimates\label{t:Andvel}} 
\tablehead{
\colhead{Method} & \colhead{\qquad\qquad\qquad $v_{\rm LOS}$} & 
\colhead{\qquad\qquad\qquad $v_W$} & 
\colhead{\qquad\qquad\qquad $v_N$} \\
& km/s & km/s & km/s \\
(1) & (2) & (3) & (4) }
\startdata
\multicolumn{4}{l}{{\it HST PMs + Internal Kinematics Model +
Viewing Perspective (Section~\ref{s:intkin})}}\\
\hline
Spheroid Field     & $\ldots$ & $-179.1 \pm  64.1$ & $-122.6 \pm  60.0$ \\
Disk Field         & $\ldots$ & $-158.0 \pm  92.4$ & $  -0.5 \pm  91.3$ \\
Stream Field       & $\ldots$ & $-126.3 \pm 103.6$ & $-247.5 \pm 102.1$ \\ 
% HSTPM/inplanerot4.out???7
{\bf Weighted Av.} & $\ldots$ & $-162.8 \pm  47.0$ & $-117.2 \pm  45.0$ \\
% HSTPM2/M31v7.out
\hline
\multicolumn{4}{l}{{\it Analysis of Satellite LOS Kinematics (Section~\ref{s:satkin})}}\\
\hline
M31 Satellites     & -279.3 $\pm$  16.4 & -176.1 $\pm$ 144.1 &  8.4 $\pm$  85.4 \\
% modelfit2_out/modelfit2.10out1
M33 PM             & -183.1 $\pm$  84.9 & -47.7 $\pm$  88.2 &  70.9 $\pm$  91.5 \\
% inplanerot3_out/inplanerot3.7out_M33
IC 10 PM           & -346.1 $\pm$  84.8 & -16.2 $\pm$  88.0 & -47.3 $\pm$  89.3 \\
% inplanerot3_out/inplanerot3.7out_IC10
Outer LG Galaxies  & -361.3 $\pm$  83.6 & -140.5 $\pm$ 58.0 &-102.6 $\pm$  52.5 \\
% baryvel.out6
{\bf Weighted Av.} & -281.1 $\pm$  15.6 &  -97.0 $\pm$ 40.7 & -45.1 $\pm$  36.6 \\
% weightedav.out7
\hline
\multicolumn{4}{l}{{\it All methods combined}}\\
\hline
{\bf Weighted Av.} & $\ldots$           & -125.2 $\pm$ 30.8 & -73.8 $\pm$  28.4 \\
% HSTPM2/M31v7.out
\enddata
\tablecomments{\small Estimates of the heliocentric velocity of the
M31 COM from different methods, as indicated in column~(1).  The top
part of the table gives the results from the HST PM measurements,
corrected for internal kinematics as described in
Section~\ref{s:intkin}. The weighted average of the results from the
three different HST fields is listed as well. The middle part of the
table gives the results from the updated vdMG08 analysis, based on the
kinematics of M31 and LG satellite galaxies, as described in
Section~\ref{s:satkin}. The weighted average of the four independent
estimates is listed as well. Column~(2) lists the estimated M31
systemic line-of-sight velocities (the actual velocity measured
directly from M31 itself is known to be $-301 \pm 1$ km/s; vdMG08).
Columns~(3) and~(4) list the estimated M31 transverse velocities in
the West and North directions, respectively. The bottom line of the
table lists the weighted average of the two weighted averages from the
different methods. This is the final result used in the remainder of
our study.}
\end{deluxetable}

%%% END TABLES %%%

%%%%%%%%%%%%%%%
% Start captions + figures on a new page
%%%%%%%%%%%%%%%

\clearpage

%%%%%%%%%%%%%%%
% Figures
%%%%%%%%%%%%%%%

\clearpage

\clearpage

\clearpage

\clearpage


\begin{thebibliography}{}

\bibitem[]{Bar11}
Barnes, J. E., 2011, Astrophysics Source Code Library, record ascl:1102.027

\bibitem[]{Bes07} 
Besla, G., Kallivayalil, N., Hernquist, L.,
Robertson, B., Cox, T. J., van der Marel, R. P., \& Alcock, C. 2007,
ApJ, 668, 949

\bibitem[]{Boy11}
Boylan-Kolchin, M., Besla, G., \& Hernquist, L. 2011, MNRAS, 414, 1560

\bibitem[]{Bra04}
Braun, R., \& Thilker, D. 2004, A\&A, 417, 421

\bibitem[]{Bro06} 
Brown, T. M., Smith, E., Guhathakurta, P., Rich, R. M., Ferguson,
H. C., Renzini, A., Sweigart, A. V., \& Kimble, R. A. 2006, ApJL, 
636, L89

\bibitem[]{Bru05}
Brunthaler, A., Reid, M. J., Falcke, H., Greenhill, L. J., \& Henkel, C.
2005, Science, 307, 1440

\bibitem[]{Bru07}
Brunthaler, A., Reid, M. J., Falcke, H., Henkel, C., \& Menten, K. M.
2007, A\&A, 462, 101

\bibitem[]{Cha06}
Chapman, S. C., Ibata, R., Lewis, G. F., Ferguson, A. M. N., Irwin, M., 
McConnachie, A., \& Tanvir, N. 2006, 653, 255

\bibitem[]{Cha07}
Chapman, S. C., et al. 2007, ApJ, 662, L79

\bibitem[]{Col09}
Collins, M. L. M., et al. 2009, MNRAS, 396, 1619 (C09)

% \bibitem[]{Col10}
% Collins, M. L. M., et al. 2010, MNRAS, 407, 2411 (C10)

\bibitem[]{Cor00}
Corbelli, E., \& Salucci, P. 2000, MNRAS, 311, 441

\bibitem[]{Cor97}
Corbelli, E. \& Schneider, S. E. 1997, ApJ, 479, 244

\bibitem[]{Cor03}
Corbelli, E. 2003, MNRAS, 342, 199

\bibitem[]{Cor10}
Corbelli, E., Lorenzoni, S., Walterbos, R., Braun, R., \& Thilker, D.
2010, A\&A, 511, 89

\bibitem[]{Cox08}
Cox, T. J., \& Loeb, A. 2008, MNRAS, 386, 461

\bibitem[]{Deh98}
Dehnen, W., Binney, J. J. 1998, MNRAS, 298, 387

\bibitem[]{Dor12}
Dorman, C., et al. 2012, ApJ, submitted

\bibitem[]{Ein82}
Einasto, J., \& Lynden-Bell, D. 1982, MNRAS, 199, 67

\bibitem[]{Eva00}
Evans, N. W., Wilkinson, M. I., Guhathakurta, P., Grebel, E. K., 
Vogt, S. S. 2000, ApJ, 540, L9

\bibitem[]{Far06}
Fardal, M. A., Babul, A., Geehan, J. J., \& Guhathakurta, P. 2006,
MNRAS, 366, 1012

\bibitem[]{Far07}
Fardal, M. A., Guhathakurta, P., Babul, A., \& McConnachie, A. W. 2007,
MNRAS, 380, 15

\bibitem[]{Far08}
Fardal, M. A., Babul, A., Guhathakurta, P., Gilbert, K., \& Dodge, C. 2008,
ApJ, 682, L33

\bibitem[]{Fer00}
Ferguson, A. M. N., Gallagher, J. S., \& Wyse, Rosemary F. G,
2000, AJ, 120, 821

\bibitem[]{Gal05}
Galleti, S., Bellazzini, M., Federici, L., \& Fusi Pecci, F. 2005,
A\&A, 436, 535 (G05)

\bibitem[]{Gal07}
Galleti, S., Bellazzini, M., Federici, L., Buzzoni, A., \& 
Fusi Pecci, F. 2007, A\&A, 471, 127 (G07)

\bibitem[]{Gee06}
Geehan, J. J., Fardal, M. A., Babul, A., \& Guhathakurta, P. 2006, MNRAS,
366, 996

\bibitem[]{Ghe08}
Ghez, A. M., et al. 2008, ApJ, 689, 1044

\bibitem[]{Gil07}
Gilbert, K. M. 2007, ApJ, 668, 245

\bibitem[]{Gilb09}
Gilbert, K. M., Guhathakurta, P., Kollipara, P., Beaton, R. L., 
Geha, M. C., Kalirai, J. S., Kirby, E. N., Majewski, S. R., 
Patterson, R. J. 2009, ApJ, 705, 1275

\bibitem[]{Gill09}
Gillessen, S., Eisenhauer, F., Trippe, S., Alexander, T., Genzel, R., 
Martins, F., \& Ott, T. 2009, ApJ, 692, 1075

\bibitem[]{Got78}
Gott, J. R., \& Thuan, T. X. 1978, ApJ, 223, 426

\bibitem[]{Guh05}
Guhathakurta, P., Ostheimer, J. C., Gilbert, K. M., Rich, R. M., 
Majewski, S. R., Kalirai, J. S., Reitzel, D. B., \& Patterson, R. J. 2005,
ArXiv e-prints, astro-ph/0502366

\bibitem[]{Guh06}
Guhathakurta, P., et al. 2006, AJ, 131, 2497

\bibitem[]{Guo10}
Guo, Qi, White, S. D. M., Li, C., \& Boylan-Kolchin, M. 2010, 
MNRAS, 404, 1111

\bibitem[]{Iba04}
Ibata, R., Chapman, S., Ferguson, A. M. N., Irwin, M., 
Lewis, G. F., \& McConnachie, A. W. 2004, MNRAS, 351, 117

\bibitem[]{Iba07}
Ibata, R., Martin, N. F., Irwin, M., Chapman, S., Ferguson, A. M. N., 
Lewis, G. F., \& McConnachie, A. W. 2007, ApJ, 671, 1591

\bibitem[]{Irw05}
Irwin, M. J., Ferguson, A. M. N., Ibata, R. A., Lewis, G. F., \& Tanvir, N. R.
2005, ApJL, 628, L105

\bibitem[]{Jar11}
Jarosik, N. et al. 2011, ApJS, 192, 14

\bibitem[]{Kah59}
Kahn, F. D., \& Woltjer, L. 1959, ApJ, 130, 705

\bibitem[]{Kal06a}
Kalirai, J. S., Guhathakurta, P., Gilbert, K. M., Reitzel, D. B., Majewski, 
S. R., Rich, R. M., Cooper, M. C., 2006a, ApJ, 641, 268

\bibitem[]{Kal06b}
Kalirai, J. S., et al. 2006b, ApJ, 648, 389

\bibitem[]{Ker86}
Kerr, F. J., \& Lynden-Bell, D. 1986, MNRAS, 221, 1023

\bibitem[]{Kly02}
Klypin, A., Zhao., H. S., \& Somerville, R. S. 2002, 573, 597

\bibitem[]{Kly11}
Klypin, A., Trujillo-Gomez, S., Primack, J. 2011, ApJ, in press 
[arXiv:1002.3660]

\bibitem[]{Koc06} 
Koch, A., \& Grebel, E. K. 2006, AJ, 131, 1405

\bibitem[]{Koc96}
Kochanek, C. S. 1996, ApJ, 457, 228

\bibitem[]{Kro91}
Kroeker, T. L., \& Carlberg, R. G. 1991, ApJ, 376, 1

\bibitem[]{Lew07}
Lewis, G. F., Ibata, R. A., Chapman, S. C., McConnachie, A., Irwin, M. J., 
Tolstoy, E., \& Tanvir, N. R. 2007, MNRAS, 375, 1364

\bibitem[]{Li08}
Li, Y.-S., \& White, S. D. M. 2008, MNRAS, 384, 1459

\bibitem[]{Loe05}
Loeb, A., Reid, M. J., Brunthaler, A., \& Falcke, H.
2005, ApJ, 633, 894

\bibitem[]{Lyn81}
Lynden-Bell, D. 1981, The Observatory, 101, 111

\bibitem[]{Lyn99}
Lynden-Bell, D. 1999, in ``The stellar content of Local Group
galaxies'', Proc.~IAU Symp.~192, Whitelock, P., \& Cannon, R., eds.,
p.~39 (San Francisco: Astronomical Society of the Pacific)

\bibitem[]{Maj07}
Majewski, S. R., et al. 2007, ApJL, 670, L9

\bibitem[]{Mar09}
Martin, N. F., et al. 2009, ApJ, 705, 758

\bibitem[]{Mat08}
Mateo, M, Olszewski, E. W., \& Walker, M. G. 2008, ApJ, 675, 201

\bibitem[]{Mac03}
McConnachie, A. W., Irwin, M. J., Ibata, R. A., Ferguson, A. M. N., Lewis,
G. F., \& Tanvir, N. 2003, MNRAS, 343, 1335

\bibitem[]{Mac04}
McConnachie, A. W., Irwin, M. J., Ferguson, A. M. N., Ibata, R. A., Lewis,
G. F., \& Tanvir, N. 2004, MNRAS, 350, 243

\bibitem[]{Mac05} 
McConnachie, A. W., Irwin, M. J., Ferguson, A. M. N., Ibata, R. A.,
Lewis, G. F., \& Tanvir, N. 2005, MNRAS, 356, 979

\bibitem[]{McC06}
McConnachie, A. W., \& Irwin, M. J. 2006, MNRAS, 365, 902

\bibitem[]{Mac08}
McConnachie, A. W. et al. 2008, ApJ, 688, 1009

\bibitem[]{Mac09}
McConnachie, A. W. et al. 2009, Nature, 461, 66

\bibitem[]{MaG05}
McGaugh, S. S. 2005, ApJ, 632, 859

\bibitem[]{McM10}
McMillan, P. J., \& Binney, J. J. 2010, MNRAS, 402, 934

\bibitem[]{McM11}
McMillan, P. J. 2011, MNRAS, 414, 2446

\bibitem[]{Mer06}
Merrett, H. R., et al. 2006, MNRAS, 369, 120

\bibitem[]{Nav97}
Navarro, J. F., Frenk, C. S., White, S. D. M. 1997, ApJ, 490, 493

\bibitem[]{Net07}
Neto, A. F., et al. 2007, MNRAS, 381, 1450

\bibitem[]{Pee01}
Peebles, P. J. E., Phelps, S. D., Shaya, E. J., \& Tully, R. B. 2001,
ApJ, 554, 104

\bibitem[]{Ray89}
Raychaudhury, S., \& Lynden-Bell, D. 1989, MNRAS, 240, 195

\bibitem[]{Rei04}
Reid, M. J., \& Brunthaler, A. 2004, ApJ, 616, 872

\bibitem[]{Rei09}
Reid, M. J., et al. 2009, ApJ, 700, 137

\bibitem[]{San86}
Sandage, A. 1986, ApJ, 307, 1

\bibitem[]{Sch10}
Sch\"nrich, R., Binney, J., \& Dehnen, W. 2010, MNRAS, 403, 1829

\bibitem[]{Sha09}
Shattow, G., \& Loeb, A., 2009, MNRAS, 392, L21

\bibitem[]{Soh12}
Sohn, S. T., Anderson, J., \& van der Marel, R. P. 2012, ApJ, 
submitted (Paper~I)

\bibitem[]{Sta01} 
Stadel, J. G. 2001, Ph.D.~thesis ``Cosmological
N-body simulations and their analysis", University of Washington

\bibitem[]{Tol12}
Tollerud, E., et al. 2012, ApJ, submitted [arXiv:1112.1067v1]

\bibitem{vdB98}
van den Bergh, S. 1998, AJ, 116, 1688

\bibitem{vdB00}
van den Bergh, S. 2000, The Galaxies of the Local Group (Cambridge: Cambridge
University Press)

\bibitem[]{vdM02}
van der Marel, R. P., Alves, D. R., Hardy, E., \& Suntzeff, N. B.
2002, AJ, 124, 2639 

\bibitem[]{vdM08}
van der Marel, R. P., \& Guhathakurta, P. 2008, ApJ, 678, 187 (vdMG08)

\bibitem[]{Wan06}
Wang, L., Li, C., Kauffmann, G., \& De Lucia, G. 2006, MNRAS, 371, 537

\bibitem[]{Wat10}
Watkins, L. L., Evans, N. W., \& An, J. H. 2010, MNRAS, 406, 264

\bibitem[]{Zar89}
Zaritsky, D., Olszewski, E., Schommer, R., Peterson, R., Aaronson, M.
1989, ApJ, 345, 759

\bibitem[]{Zar94}
Zaritsky, D., \& White, S. D. M. 1994, ApJ, 435, 599

\end{thebibliography}
\end{document}